\documentclass[aps,pre,twocolumn,showpacs,amsmath,amssymb,floatfix]{revtex4}
\usepackage{graphicx,color, bm} 
\usepackage{dcolumn} 

\newcommand{\be}{\begin{equation}}
\newcommand{\ee}{\end{equation}}

\newcommand{\bomega}{\mbox{\boldmath{$\omega$}}}

\begin{document}

\title{Ideal evolution of MHD turbulence when imposing Taylor-Green symmetries}

\author{M.E. Brachet$^1$, M. D. Bustamante$^2$, G. Krstulovic$^3$, P.D. Mininni$^{4,5}$, A. Pouquet$^4$  and D. Rosenberg$^4$}
\affiliation{
$^1$ \ Laboratoire de Physique Statistique de l'\'Ecole Normale Sup\'erieure,
associ\'e au CNRS et aux Universit\'es ParisVI et VII, 24 Rue Lhomond, 75231 Paris, France. \\
$^2$\
{School of Mathematical Sciences,
                University College Dublin,
                Belfield, Dublin 4, Ireland.}
\\
$^3$\ {Laboratoire Lagrange UMR7293, Universit\'e de Nice Sophia-Antipolis, CNRS, Observatoire de la C\^ote d'Azur, B.P. 4229, 06304 Nice Cedex 4, France}\\
$^4$\ Computational and Information Systems Laboratory, NCAR, P.O. Box 3000, Boulder CO 80307, USA.\\
$^5$\ Departamento de F\'\i sica, Facultad de Ciencias Exactas y Naturales, Universidad de Buenos Aires and IFIBA, CONICET,
                 Ciudad Universitaria, 1428 Buenos Aires, Argentina.}


\begin{abstract}
We investigate the ideal and incompressible magnetohydrodynamic (MHD) equations in three space dimensions for the development  of potentially singular structures. The methodology consists in implementing the four-fold symmetries of the Taylor-Green vortex generalized to MHD, leading to substantial computer time and memory savings at a given resolution; we also use a re-gridding method that allows for lower-resolution runs at early times, with no loss of spectral accuracy. One magnetic configuration is examined at an equivalent resolution of $6144^3$ points, and three different configurations on grids of $4096^3$ points. At the highest resolution, two different current and vorticity sheet systems are found to collide, producing two successive accelerations in the development of small scales. At the latest time, a  convergence of magnetic field lines to the location of maximum current is probably leading locally to a  strong bending and directional variability of such lines. A novel analytical method, based on sharp analysis inequalities, is used to assess the validity of the finite-time singularity scenario. This method allows one to rule out spurious singularities by evaluating the rate at which the logarithmic decrement of the analyticity-strip method goes to zero. The result is that the finite-time singularity scenario cannot be ruled out, and the singularity time could be somewhere between $t=2.33$ and $t=2.70.$ More robust conclusions will require higher resolution runs and grid-point interpolation measurements of maximum current and vorticity.
\end{abstract}
\maketitle

\section{Introduction}\label{s:intro}

The class of problems addressing the formation of singularities and the existence and structure of solutions of nonlinear partial differential equations for all times forms an important branch of mathematics, with wide application in numerous fields: engineering, astro-- and geophysics, laboratory studies of superfluids, and in meteorological research on extreme events such as tornadoes and hurricanes. The presence or absence of either dissipation-viscosity, of magnetic resistivity in magnetohydrodynamics (MHD), or of dispersion, plays an essential role as well. The significance of such questions is recognized for example by the Clay Institute Millennium Prize for a proof of existence and smoothness of finite energy solutions of the Navier-Stokes equations, and by the numerous studies devoted to them: How fast do (potentially) singular structures form? What is their temporal evolution and geometry? What role do their interactions play and how might they lead to a modification of transport properties within complex flows, including in the presence of magnetic fields? Progress on such problems will most likely come from a combination of mathematics, laboratory experiments and direct numerical simulations (DNS), in the latter case in particular using computer codes with high accuracy and performing studies at the highest possible resolutions.

There is a large body of analytical and numerical work on singularities in fluids. As theoretical estimates are not necessarily sharp, numerical data are invaluable in assessing potential singularities, as discussed, e.g., in \cite{constantin1995}. Unfortunately, in the case of the numerics, with regard to existence the answer vacillates between ``yes'' and ``no'' \cite{eyink_euler}. Singularities occur in simplified models, as derived in \cite{vieille_82, vieille_84} assuming an isotropic pressure Hessian. These models have been generalized to MHD in the vicinity of magnetic null points \cite{klapper_96} and lead as well to a singularity, but the question remains open in the general (and most physically relevant) case.

One of the most useful criteria in the search for a singularity comes from the Beale-Kato-Majda theorem (BKM hereafter) \cite{BKM}
which states for incompressible ideal fluids that, if the flow presents a finite-time singularity at $T_{\ast}$, then
\begin{equation}
\int_0^{T_{\ast}}||{\bf {\bomega}}(.,t)||_{\infty} dt= \infty,
\label{BKM} \end{equation}
where we have used the usual notation for the ${\cal L}_{\infty}$ or supremum norm, ${\bomega}=\nabla \times {\bf v}$ being the vorticity and ${\bf v}$ the velocity. If a power-law divergence of vorticity at $T_\ast$ is assumed, $||{\bf {\bomega}}(.,t)||_{\infty} = C |T_{\ast}-t|^{-\beta},$ for $t \to T_\ast,$ with $\beta > 0$ and $C$ a constant, then the BKM theorem can be re-expressed as: ``The flow has a finite-time singularity at $t=T_\ast$ if and only if $\beta \geq 1.$'' As stressed in \cite{kerr_05}, enstrophy production $D\Omega/Dt$, with $\Omega= \int {\bomega}^2({\bf x})d^3{\bf x}$ (the ${\cal L}_2$ norm), should also be monitored to detect singularities, and care must be taken in assessing $C, \beta$ and $T_{\ast}$ when fitting the data stemming from DNS.

Furthermore, the dynamics of the vorticity (and current in MHD flows) should be monitored not only for their ${\cal L}_2$
and ${\cal L}_{\infty}$ norms, but also for changes in the direction of their field lines (or ``swing''  \cite{gibbon_08}).
It was found in neutral flows that the rapid growth of $||{\bf \bomega}||$ can be countered by the straightening of
vortex lines (see \cite{branden1995, pouquet_sanminiato_96,gibbon_08} in the MHD case). Moreover, the study of
the evolution of the curvature and torsion of vortex (or current) lines yields interesting insights into the dynamics of
ideal flows \cite{constantin1995}. The rich variety in the observed behavior has resulted in a plethora of initial conditions
examined in previous numerical studies.

Among the 3D flows that have been considered for their potential singular behavior for ideal (non-dissipative) fluids are the Taylor-Green flow (TG hereafter)  \cite{brachet_83}, the Kida-Pelz flow (KP) \cite{kida_85, boratav_94}, and two anti-parallel vortices  \cite{kerr_89, kerr_93}, all displaying symmetries that can be implemented numerically (see also \cite{pelz_01}). These flows have been studied by several teams, with a recent revival \cite{kerr_05, hou_06, hou_08, kerr_08, bustamante} (see, e.g., \cite{eyink_euler} for a brief introduction to the literature).

In MHD when coupling to a magnetic field, the  theorem equivalent to BKM involves the sum of the maxima of vorticity
${\mathbf \bomega}$ and current density \cite{caflisch}.  Ideal MHD in two space dimensions  has been studied in the past (see \cite{uf_OT2D, klapper_93, grauer_97, grauer_98a, klapper_98} and more recently using high-resolution runs \cite{krstu_11}), but in the 3D general case, ideal runs are scarce except for the pioneering work using symmetric configurations of linked flux tubes with zero initial velocity \cite{kerr_branden_99}, or with adaptive mesh refinement (AMR) using
finite differences \cite{grauer_98b,  grauer_00, grafke}.

One can also use the TG flow and generalize it to MHD (hereafter, TG-MHD flows), as done in \cite{lee_ideal}. One of the TG-MHD flows studied for its possible singular behavior in \cite{lee_ideal} displays a feature not observed at the time in the fluid case: after an initial phase of thinning of the current and vortex sheets, the flow outside the structure pushes together two current sheets with widely different directions of the magnetic field embedded in them, leading to a rotational quasi-discontinuity with a substantial acceleration in the development of small scales.
Once dissipation is restored, this small-scale activity is diagnosed as intermittent reconnection \cite{lee2}. Rotational and tangential discontinuities, identified as intermittent structures, have been observed in the Solar Wind using a variety of in-situ acquired data \cite{veltri1999}; they have also been identified at the edge of Reverse Field Pinch plasma devices (see, e.g., \cite{veltri2009} for review).
Using the Cluster ensemble of four satellites, all four spacecrafts indicate at times a directional (either rotational or tangential)
discontinuity, including with a small normal component of the magnetic field.
Such rotational discontinuities can stem from non-linear steepening or from reconnection of magnetic field lines \cite{Lin2009}. Their modeling leads to statistical properties akin to that of so-called nano-flares observed in the solar corona \cite{veltri2005}, and they provide tantalizing hints that singularities may exist in MHD.

Only by performing substantially higher-resolution and high accuracy runs that high-performance computing resources can allow, shall we be able to explore several configurations leading to possible singular behavior in MHD.
It is in this context that we propose to search in this work for singularities in MHD with different configurations and using the highest known resolutions (and hence, scale separation between the size of the box and the size of the mesh); thus, a run is performed on an equivalent grid of $6144^3$ points in one case, following on the work done in \cite{lee_ideal} on grids of $2048^3$ points.

\section{Numerical procedure} \label{s:proc}
In this section we describe briefly the codes and methods used, and give details of the numerical simulations. We present first the MHD equations, and then introduce the initial conditions.  Then, we explain how the code is parallelized for the simulations at the largest resolutions. The choice of de-aliasing method is crucial to conserve the total energy and other quadratic invariants with good accuracy, and details concerning our methodology are given and compared with other choices for de-aliasing. Then, the procedure followed to increase spatial resolution as structures become thinner is explained.
Finally, we comment on the effect that imposing the four-fold symmetries of the Taylor-Green vortex generalized to MHD might have.

\subsection{Equations, initial conditions and the TYGRS code} \label{ss:eq}

The MHD equations for an incompressible and ideal fluid with $\bf {v}$ and $\bf {b}$ respectively the velocity and magnetic field read:
\begin{eqnarray}
&& \frac{\partial {\bf v}}{\partial t} + {\bf v} \cdot \nabla {\bf v} =
    -\frac{1}{\rho_0} \nabla {\cal P} + {\bf j} \times {\bf b}  \ ,
    \label{eq:MHDv} \\
&& \frac{\partial {\bf b}}{\partial t} = \nabla \times ( {\bf v} \times {\bf b}) \  ;
\label{eq:MHDb}
\end{eqnarray}
$\rho_0=1$ is the (uniform) density and ${\bf b}$ is the Alfv\'en velocity, ${\cal P}$ is the pressure,
${\bf \nabla} \cdot {\bf v} = 0 \ , \nabla \cdot {\bf b} = 0$, and there are no dissipative or forcing terms; finally, ${\bf j}=\nabla \times {\bf b}$ is the current density.
The total  (kinetic plus magnetic) energy $E_T$, the cross helicity $H_C$ and the magnetic helicity $H_M$, defined as
\begin{eqnarray}
E_T=E_V+E_M&=&\left<v^2+b^2\right>/2 \\
H_C=\left<{\bf v} \cdot {\bf b}\right> &,& H_M=\left<{\bf A} \cdot {\bf b}\right>
\end{eqnarray}
with ${\bf A}$ the magnetic potential (${\bf b} = \nabla \times {\bf A}$), are all conserved by the nonlinear interactions \cite{woltjer}.

In practice, a pseudo-spectral code solves these equations in Fourier space, truncated up to some maximum wavenumber. The truncated MHD equations for the Fourier modes ${\bf u}_{\bf k}$ and ${\bf b}_{\bf k}$, with $k\in [k_{\rm min}, k_{\rm max})$ can be written easily, the Fourier modes satisfying ${\bf u}_{\bf k}=0, {\bf b}_{\bf k}=0$ if $|{\bf k}|\ge k_{\rm max}$ or if $|{\bf k}|< k_{\rm min}$. For a computational box of length $2\pi$, we have $k_{\rm min}=1$, and with a de-aliasing using the 2/3-rule, $k_{\rm max}=N/3$, where $N$ is the number of modes per dimension (we assume a box with unit aspect ratio). Other de-aliasing methods can be used successfully \cite{hou_06, hou_08} and are discussed briefly below (see Sec.~\ref{ss:trunc}). It is important to note here that de-aliasing is crucial in pseudo-spectral simulations to remove spurious growth of modes with large wavenumbers, and to conserve the total energy and other quadratic invariants.
Indeed, a pseudo-spectral code which is fully dealiased is equivalent to a Galerkin truncation, and thus preserves all quadratic invariants in the system to round-off error.

The equations are solved starting from initial conditions for the velocity and the magnetic field. If the initial conditions have symmetries that are preserved by the equations, then the symmetries can be used to save memory and computing time. As already mentioned, in hydrodynamics (${\bf b} \equiv 0$) one of the simplest velocity fields satisfying the symmetries of the equations is the TG flow (note that the $z$ component, initially equal to zero, will grow with time) \cite{brachet_83, brachet_92, cicho_05a}:
\begin{equation}
\bm u(x, y, z) = u_0 \left[ (\sin x \cos y \cos z) \bm{\hat{e}_x} - (\cos x \sin y \cos z)\bm{\hat{e}_y} \right]   \ . \label{eqn:simple_vTG}  \end{equation}
It is interesting to point out that the TG flow in a periodic box shares similarities with the von K\`arm\`an flow between two counter-rotating disks as used in several laboratory experiments, including those with liquid metals such as sodium or gallium, to study the generation of magnetic fields.

To generalize the TG flow to MHD, we use the velocity as prescribed by Eq.~(\ref{eqn:simple_vTG}), and we will consider three possible choices for the initial magnetic field $\bf {b}$ with the same overall symmetries \cite{lee_ideal, lee2, gafd}. We refer to these three flows as the insulating (I) defined by
\begin{equation}
{\bf b^i}=b_0^i \begin{pmatrix}
 \cos x\sin y\sin z \\
 \sin x\cos y\sin z  \\
-2 \sin x\sin y\cos z
 \end{pmatrix},\label{eqn:btg_I}
\end{equation}
the \emph{alternative} insulating flow (A):
\begin{equation}
{\bf b^a}=b_0^a \begin{pmatrix}
\cos2x\sin2y\sin2z \\
-\sin2x\cos2y\sin2z  \\
0
 \end{pmatrix},
\end{equation}
and the conducting flow (C):
\begin{equation}
{\bf b^c}=b_0^c \begin{pmatrix}
\sin2x\cos2y\cos2z \\
\cos2x\sin2y\cos2z  \\
-2 \cos2x\cos2y\sin2z
 \end{pmatrix}. \label{eqn:btg_C}
\end{equation}
Note that the I, A and C flows, with almost identical invariants, have nevertheless three different developed energy spectra in the non-ideal case at the maximum of dissipation \cite{lee2}, displaying a lack of universality in MHD turbulence in the absence of an imposed magnetic field.

For all three configurations, $E_V=E_M=0.125$ when $u_0=1$ and $b_0=\sqrt{1/3}$, 1, $\sqrt{2/3}$, respectively; for the helicities, $H_M\equiv 0$ and $H_C\sim 0$ because of the imposed symmetries; note however that there can be strong local correlations corresponding to local alignment of ${\bf u}$ and ${\bf b}$, as can be shown both analytically and numerically  \cite{matthaeus_08}. In the I case, the current $\bm j^i$ is everywhere parallel to the walls of the so-called impermeable box $[0,\pi]^3$ which thus appears to be insulating. For the C case, ${\bf j}^c$ in the $[0,\pi]^3$ box is perpendicular to the walls, which are therefore conducting. In this configuration, $H_C$ is non-zero but small (less than $4\%$ at its maximum over time, in a dimensionless measure relative to the total energy). Finally, $\bm b^a$ is an alternative insulating MHD vortex.

The code, TYGRS (TaYlor-GReen Symmetric; see below), enforces the symmetries of the TG vortex in 3D hydrodynamics, and of the TG-MHD vortices in 3D MHD within the periodic cube of length $2\pi$. These symmetries include: mirror symmetries about the planes
$x=0\ \&\ \pi$, $y=0\ \&\ \pi$ and $z=0\ \&\ \pi$ together with
$ \ x=\pi, \ y=0, \ y=\pi, \ z=0,$ and $z=\pi$
(e.g., in the $x$-direction: $\bm v_x(-x, y, z) = -\bm v_x(x,y,z)$ and $\bm v_x(\pi-x, y, z) = -\bm v_x(\pi+x,y,z)$),
rotational symmetries of angle $n\pi$ about the axes $(x,y,z)=(\frac{\pi}{2},y,\frac{\pi}{2})$ and $(x,\frac{\pi}{2},\frac{\pi}{2})$,
and rotational symmetries of angle $n\pi/2$ about the axis $(\frac{\pi}{2},\frac{\pi}{2},z)$, for $n \in Z$. Because of these symmetries, the Fourier-transformed fields are non-zero only for wavenumbers $({\bf k}_x,{\bf k}_y,{\bf k}_z)$ with jointly even or jointly odd components.

Thus, TYGRS computations at a given scale separation (defined as the ratio $k_{max}/k_{min}$, which is proportional to the Reynolds number in the dissipative case),  or at a given equivalent resolution, are performed on linear grids that are one-fourth the size of those for a general code, by exploiting symmetries of the TG vortex: one obtains the flow in the full periodic box of size $[0,2\pi]^3$ by applying these symmetries to the impermeable box $[0,\pi]^3$. The nonlinear terms and their temporal derivatives are computed from the even-odd decomposition of the fields in the fundamental box $[0,\pi/2]^3$.
Note that TYGRS  performs a DNS, since no modeling of small scales is done. For time integration, an explicit  $2^{nd}$-order Runge-Kutta scheme is used. Because the time integration truncation error at the proposed resolutions may exceed the single floating point precision, we use double precision for the computations.

No uniform external field ${\bf B}_0$ will be imposed in our simulations. Such an external field is known to slow-down small-scale development, and may quench the development of singularities \cite{bardos_82, uf_OT2D} because of the semi-dispersive nature of the problem, with Alfv\'en waves propagating in opposite directions along ${\bf B}_0$. This slowing-down of nonlinear dynamics due to waves has been modeled phenomenologically in several ways, starting with Iroshnikov and Kraichnan in the mid sixties with a $k^{-3/2}$ total isotropic energy spectrum, as opposed to the classical Kolmogorov spectrum for fluid turbulence. It can be evaluated analytically using weak turbulence theory for large ${\bf B}_0$ \cite{galtier_00, galtier_02}, leading to a steeper and anisotropic spectrum $\sim k_{\perp}^{-2}$, with $k_{\perp}$ referring to the direction perpendicular to ${\bf B}_0$.

\begin{figure}
 \begin{center}
 \includegraphics[width=1\columnwidth]{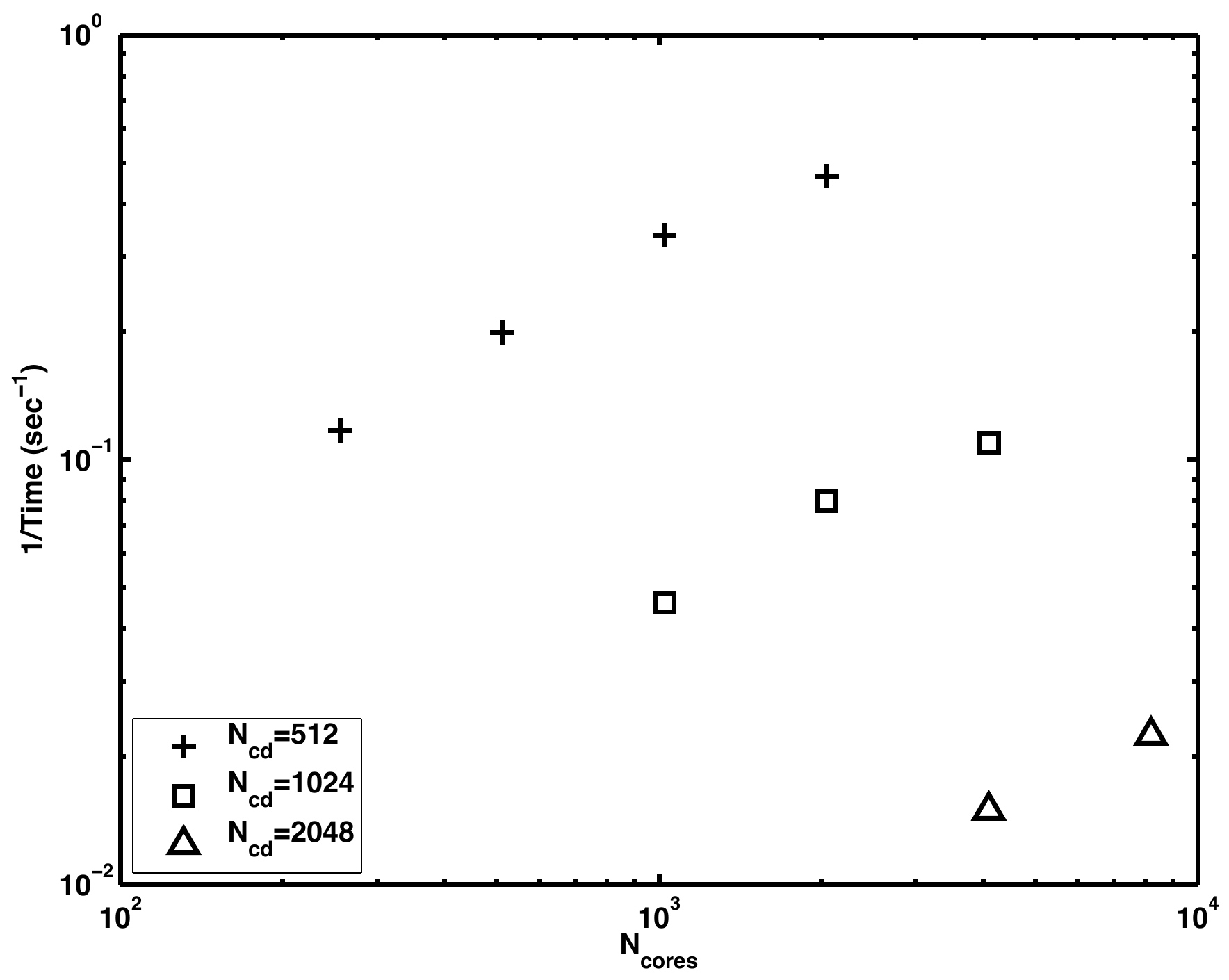}
 \caption{
Timings of the TYGRS code on Jaguar up to $\sim 10^4$ cores and up to an actual number of
computational degrees of freedom of $N_{cd}^3 = 2048^3$ points.
} \label{fig:scaling}
 \end{center}
 \end{figure}

\subsection{The role of symmetries}

In \cite{lee_ideal}, simulations with and without imposed symmetries with Taylor-Green initial conditions were compared. No differences were observed except at the lowest mode and at an energetic level close to round-off error. Also, visualization analyses showed that the physical structures that are present in the flow appear identical between the runs with and without imposed symmetries (see, e.g., \cite{gafd}). Of course, at late times instabilities develop induced by noise due to accumulated errors because, e.g., of insufficient numerical accuracy \cite{julia}. These errors can break the symmetries in the computation of the flow and field, when one does not impose the symmetries of the initial conditions. In that case, magnetic and cross-helicity grow and may lead the flow to another final state. However, this bifurcation in behavior happens at a significantly later time than the times considered in the present study.

\subsection{Implementation of the hybrid scheme for the TYGRS code} \label{ss:hybrid}

Pseudo-spectral codes are known to be optimal on  periodic domains \cite{gottlieb}. However, they require global spectral transforms, and thus are hard to implement in distributed memory environments, a crucial limitation until one-dimensional
domain decomposition techniques (DDT) arose, that allowed computation of serial Fast Fourier Transforms (FFTs) in different directions in space (local in memory) after performing transpositions. However, distributed parallelization using the Message Passing Interface (MPI) in pseudo-spectral codes is limited in the number of processors that can be used, unless more transpositions are done per FFT (thus increasing communication). The hybrid (MPI-OpenMP) scheme we have implemented for a general code builds upon a one-dimensional (slab-based) domain decomposition that is effective for parallel scaling using MPI alone \cite{hybrid}. In the scheme, each MPI task creates multiple threads using OpenMP. This method has been extended in TYGRS to the sine (cosine) with even (odd) wavenumber FFTs needed to implement the symmetries of TG flows, using loop-level OpenMP directives and multi-threaded FFTs.

The resulting quasi-linear scaling up to $\sim 10^4$ cores for TYGRS, particularly at high resolution, is displayed in Fig. \ref{fig:scaling}. The hybrid scheme implemented in TYGRS was derived from the method developed \cite{hybrid} for a similar pseudo-spectral code--Geophysical High-Order Suite for Turbulence (GHOST)--in which no symmetries are enforced, that now shows linear scaling up to more than 98,000 processors on grids of up to $8192^3$ points.

We note that the hybrid scheme used here is not the only way in which to decompose the pseudo-spectral grid. An alternative is to retain a pure MPI model \cite{yeung} in which the domain decomposition takes the form of ``pencils'' and yields a two-dimensional domain decomposition among MPI tasks, where OpenMP is not required. This technique is also found to scale well to large core counts, although large fluctuations in performance are observed even within a given processor-domain mapping. The hybrid method offers a two-level parallelization that may be more effective in mapping the domain to the hierarchical architectures that are now emerging, and better suited for environments with multiple cores per socket. The hybrid scheme may also aid in MPI memory problems, in that fewer MPI tasks require less buffer memory. This is related to the fact that, by reducing the number of MPI processes using threads, we reduce not only the number of MPI calls, but also the amount of data that must be communicated, and hence the size of the MPI buffers required. Finally, this also allows us to use parallel MPI I/O in
environments with tens of thousands of cores, as the number of MPI tasks is only a fraction of the total number of cores used.

\subsection{Choice of truncation at high wavenumber and the issue of accuracy} \label{ss:trunc}

As explained before, one issue to resolve is how best to perform the removal of spurious modes with high wavenumber, either via a de-aliasing technique using the standard 2/3-rule whereby modes are truncated at $2k_{max}/3$ where $k_{max}=N/2$ is the maximum wavenumber of the computation on a cubic grid with $N$ points on the side, or by multiplying the r.h.s.~of the evolution equations with a high-order exponential smoothing function $\rho (k) = \exp[-m_1(2k/N)^{m_2}]$, as proposed in \cite{kerr_05, hou_06, kerr_08, hou_08} with  $m_1=m_2=36$.
Using the latter method, more Fourier modes are retained in the computation, leading to an enhanced scale separation with which smaller scales can be reached for a given grid in an ideal flow, and thus the computations can in principle be performed for a longer time.

\begin{figure}
 \begin{center}
 \includegraphics[width=1\columnwidth]{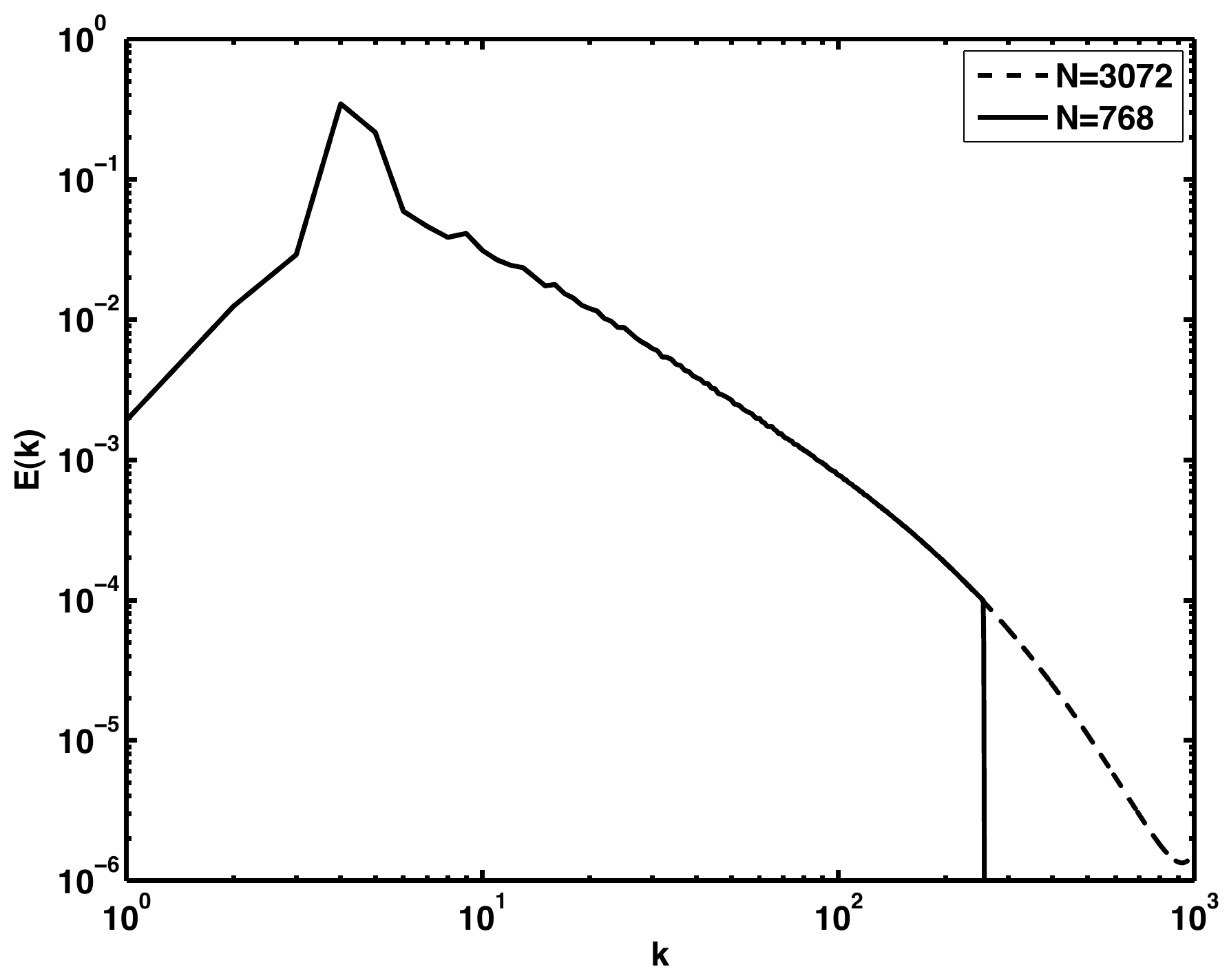}
 \caption{
 Energy spectrum for a hydrodynamic run on a grid of 3072$^3$ points, together with a N=768$^3$ truncation of that dataset using the bootstrap re-gridding down-sizing tool. The two spectra are identical (to within machine round-off) up to the truncated grid's maximum wavelength, $k_{max}=256$.
} \label{fig:spec_grid}
 \end{center}
 \end{figure}

However, when using the second method the {\emph {exact}} energy conservation in the computations is lost, as can be observed in Table \ref{tab:ener}.
Also note that the BKM criterion given in Eq.~(\ref{BKM}) for a singularity to occur is based on the supremum norm, which is more sensitive to global numerical accuracy (truncation) \cite{gaspar, NJP} and numerical precision than the ${\cal L}_2$ measures.
Furthermore, it is straightforward to check that exponential smoothing spoils the Galilean invariance ${\bf v}({\bf x},t) \to {\bf v}({\bf x}+{\bf U} t,t)-{\bf U}$ in the hydrodynamic case.
Because of these drawbacks, the 2/3 de-aliasing rule is used in the following, either in the form of the spherical rule (truncation for $|{\bf k}|\ge N/3$) or cubic rule (truncation for $|{\bf k}_x|\ge N/3$ or $|{\bf k}_y|\ge N/3$ or $|{\bf k}_z|\ge N/3$), as discussed below.

  \begin{table}[h] \begin{tabular}{|  c | c | c |}
  \hline
Time  \ & Exponential smoothing   \ &  $2/3$-cubic \\
  \hline
    \hline
  $3$ & $5.3905 \times 10^{-8}$ & $3.80216\times10^{-9}$ \\
  \hline
  $3.2$ & $6.79887\times10^{-8}$ & $4.26327\times10^{-9}$ \\
  \hline
  $3.4$ &$7.54079\times10^{-8}$ & $4.91463\times10^{-9}$ \\
  \hline
  $3.6$ & $5.79898\times10^{-8}$ & $5.81861\times10^{-9}$\\
    \hline
  $3.8$ & $-1.10333\times10^{-7}$ & $7.03868\times10^{-9}$\\
  \hline
  $4$ &  $-1.46095\times10^{-6}$ & $8.63671\times10^{-9}$\\
    \hline
   \end{tabular}   \caption{
     Time evolution of the relative error on energy conservation $\Delta E /E$ for the 
${\bf b}=0$ (hydrodynamic) Taylor Green initial data at resolution $512^3$, for two
types of spectral truncation. The ``Exponential smoothing'' method is described in
the text, while the ``$2/3$-cubic'' represents the $2/3$ de-aliasing rule using cubic 
truncation of Fourier space.}
     \label{tab:ener}   \end{table}

\subsection{The concept of bootstrap re-gridding} \label{ss:regrid}

Besides the constraints given by time stepping errors, from previous experience we know that to preserve accuracy in the computation of spatial derivatives we also need to use double precision arithmetic for a grid size at or above $4096^3$ points. On the other hand, we also know that the smallest grid size is only reached slowly (exponentially in time as long as singularities do not develop). So we propose the following question: Do we need to compute from $t=0$ to the final time at the maximum resolution $N$ that is eventually going to be needed? Indeed, at a given linear resolution $N_1<N$, one can compute until $t_{N_1}$ with sufficient accuracy, as measured for example by the logarithmic decrement technique (see below). Then, one can restart the run at $T_{N_1}$ and compute until $T_{N_2}$ with a grid of size $N_2$, with, say, $N_2=2 N_1$ grid points (not necessarily a factor of 2 of course), and this process can be re-iterated ($m$ times altogether) until we reach the desired resolution $N=2^m N_1$, so that only the last fraction of the run is done on the largest grid at the highest computational cost (in terms of both memory and CPU).

The implementation of the procedure described above requires some care when restarts are performed, from the point of view of code development because of parallelization of FFTs on grids of different sizes, as well as careful checking for accuracy for all norms, e.g., ${\cal L}_2$ but also ${\cal  L}_{\infty}$, as needed for singularity tests. However, this ``bootstrap re-gridding'' scheme allows one to save a significant fraction of compute time when carrying out the time integration at the highest resolution. In the simulations presented here, one can estimate a total cost of $1/3$ compared to the full resolution run starting at $t=0$.

It is worth pointing out that the re-gridding scheme can also be used to study the dissipative case if one chooses to start the run with the last reliable time of the ideal run. For forced runs, the extension of the methodology is straightforward. But while it may not bring about large savings, it might also  be useful in cases when the flow displays strong signs of intermittent bursts followed by long quiescent periods, as for example in the case of the stable (nocturnal) planetary boundary layer \cite{sun}, with the turbulence being related to the presence of jets at low altitude.

In practice, the re-gridding scheme takes a restart dataset in physical space, converts to wave-space, and then either truncates to reduce resolution (down-sizing, useful when performing comparisons with large eddy simulation runs), or else pads (with zeroes) in wave-space to increase the spatial resolution. The final step requires an inverse multidimensional transform at the new spectral resolution in order to convert back to physical space at the new resolution, so that the data can be used to ``restart'' at the next resolution. The end result of the equivalent down-sizing operation is illustrated in Fig. \ref{fig:spec_grid}.

\section{The I configuration at high resolution} \label{s:6000}
\subsection{Implementation of bootstrapping up to an equivalent grid of $6144^3$ points for ideal MHD} \label{ss:regrid_6144}

\begin{figure}
 \begin{center}
   \includegraphics[width=1\columnwidth]{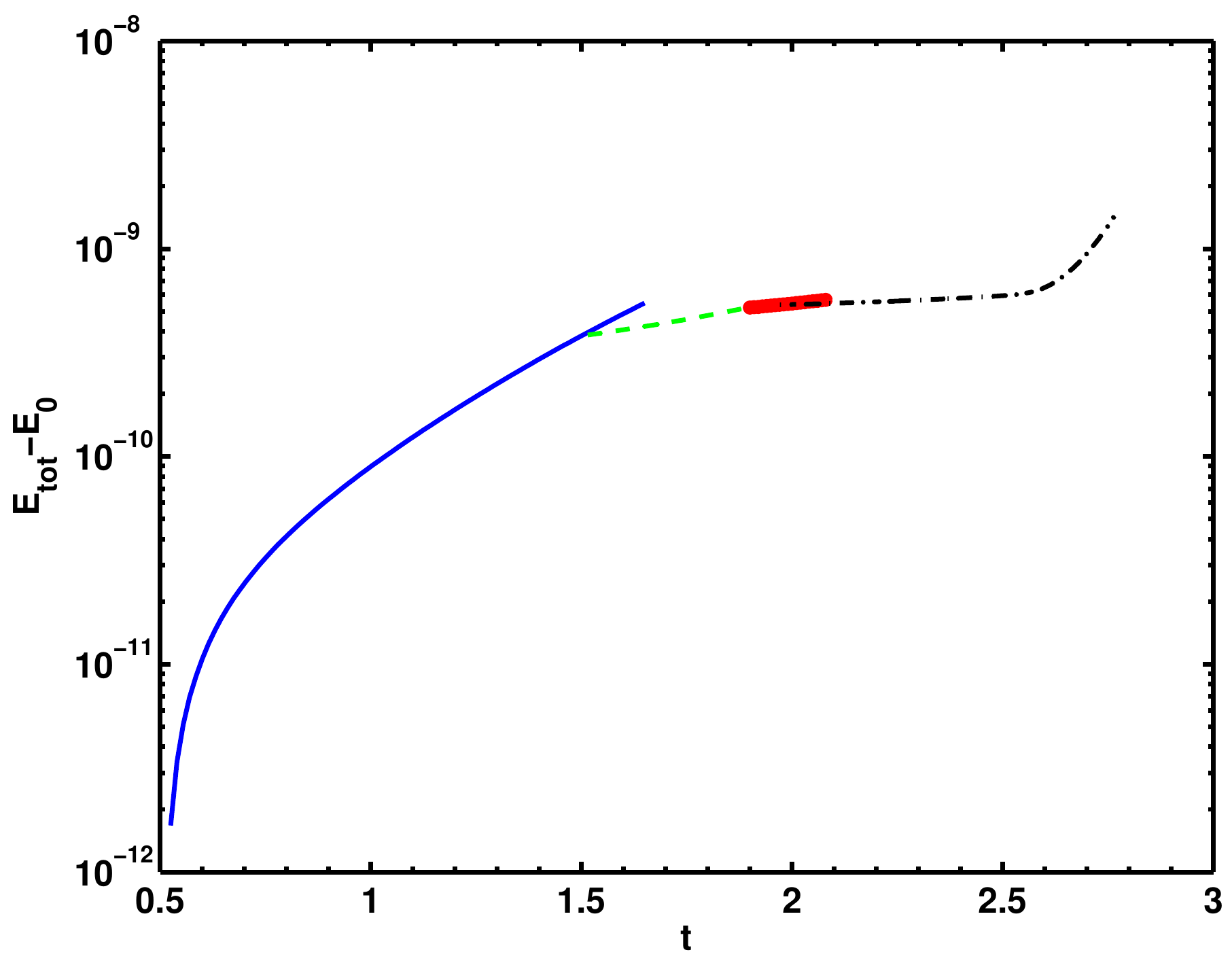}
   \includegraphics[width=1\columnwidth]{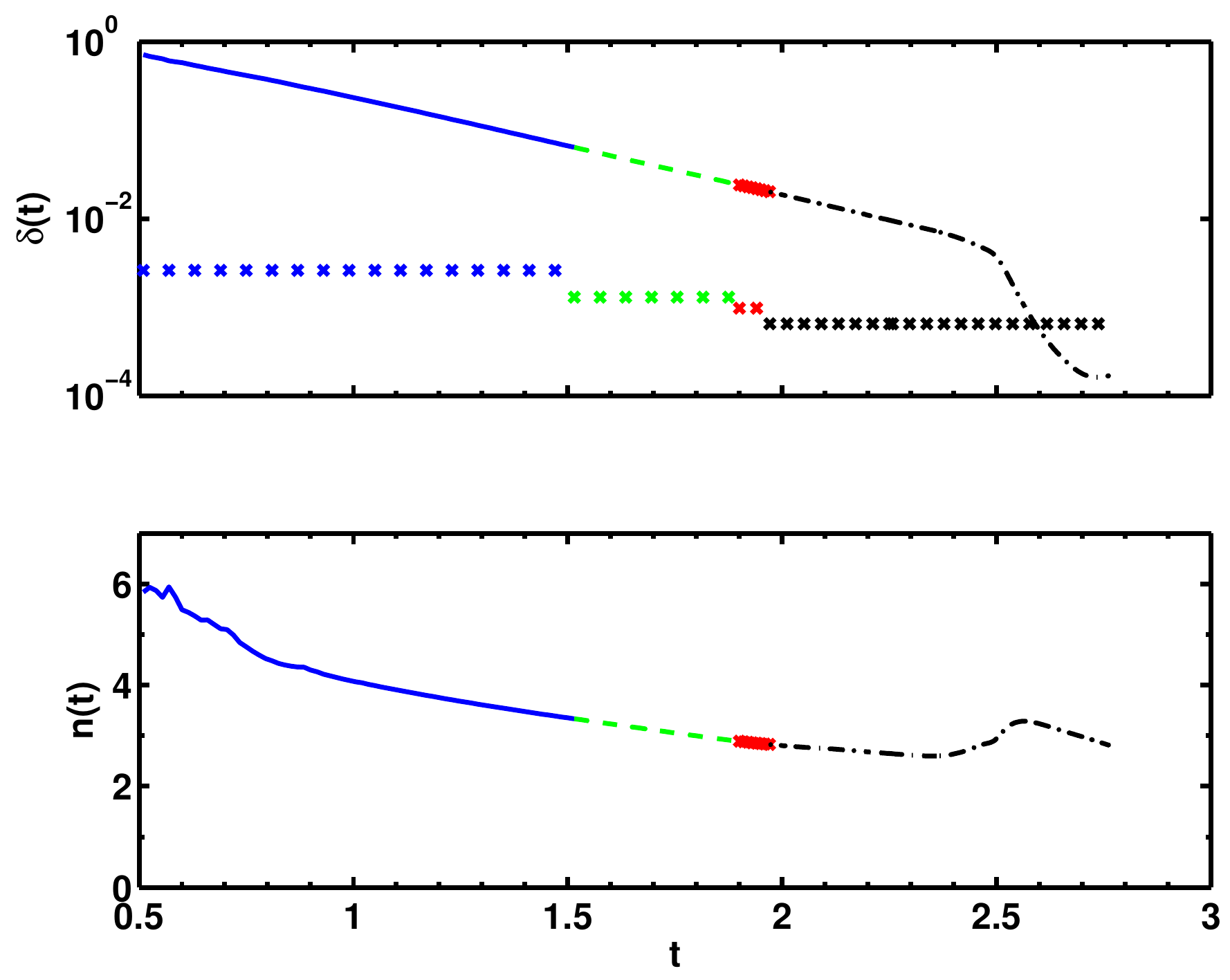}
 \caption{ {\it Top:}
  Normalized energy difference $[E_T(t)-E_0]/E_0$, with $E_0$ the initial total energy,
 showing the error growth with time. Note that most runs at a given resolution have been pursued for times longer than the time at which re-gridding was performed. Error growth is clearly slowed down by accuracy (with smaller grid resolution and smaller time-step at later times). Different colors (colors online) are used for different grid sizes, N$_i$ points per direction with N$_i$
taking values: 1536 (blue, solid); 3072 (green, dashed); 4096 (red, crosses); and 6144 (black; dash-doted).
{\it Middle and bottom:}
Temporal evolution of the logarithmic decrement $\delta$ and spectral index $n$ (bottom; see Eq.~\ref{analytic}) for the total energy spectrum (wavenumber fit interval: $[10, 1000]$). The horizontal line of crosses indicate the grid resolution limit 4/N$_i$ for a given computation on a given grid $G_i$ at an equivalent resolution of N$_i$. The color and line types are the same as in the top figure.
 }  \label{fig:spec_grid2}
 \end{center}
 \end{figure}

The bootstrapping procedure just described can in principle introduce errors in the computational procedure that breaks the spectral accuracy of the code; hence, we show now that this is not the case, provided one is careful enough in choosing the time at which the grid resolution is increased. In Fig.~\ref{fig:spec_grid2} (top) is given the normalized total energy difference (i.e., with respect to initial energy) as a function of time, with most of the error occurring at early times since the time-step is adapted to the grid spacing, which is larger earlier in the computation; the different colors (line types) indicate different grid resolutions. The energy difference remains lower than $10^{-9}$ at all times but shows a rapid increase at the latest times, indicative of a build-up of errors.
When examining the total energy spectra for different times, computed on different grid resolutions, one can observe a smooth transition from one grid to the next (not shown). It is important to note that the re-gridding is performed when the energy spectrum at the largest wavenumber in the simulation with the grid $N_i$ reaches the machine round-off level, with a cut-off conservatively chosen to be $10^{-30}$ in order to preserve a high level of accuracy throughout the run.

Apart from following the numerical conservation of the invariants of Eqs.~(\ref{eq:MHDv}) and (\ref{eq:MHDb}), with special focus on the total energy, one diagnostic has been traditionally to monitor the logarithmic decrement $\delta$, when fitting the Fourier spectrum as
\begin{equation}
E_X(k,t)=c_X(t) k^{-n_X(t)} e{^{-[2\delta_X(t)k]}}  \ ,
\label{analytic}\end{equation}
where X  stands for either the kinetic (X=V), magnetic (X=M) or total (X=T) energy, or the energies of the Els\"asser variables $E_{\pm}$ for the fields ${\bf z}_{\pm}= {\bf v}\pm {\bf b}$. As long as $\delta_X \not= 0$ the fields remain regular, and when $\delta_X$ becomes comparable to the mesh the computation of the behavior of the partial differential equations (\ref{eq:MHDv}) and (\ref{eq:MHDb}) stops, since at later times one enters the regime of statistical equilibrium. The logarithmic decrement $\delta_X$ refers to the width of the analyticity strip in the complex plane: as long as the complex singularities do not reach the real axis, the computation remains regular \cite{bardos_82, bardos_07}.
Figure \ref{fig:spec_grid2} also gives the temporal evolution of the logarithmic decrement $\delta_T$ (middle) and of the spectral index $n_T$ (bottom) for the total energy spectrum; grid resolution is indicated by the horizontal line of crosses. The fit to the spectrum (see Eq.~(\ref{analytic}) above) is done in the Fourier interval $[10,1000]$.
The acceleration in the decrease of the logarithmic decrement found in \cite{lee_ideal} is confirmed by the present computation; it is accompanied by a sharp increase in the inertial index $n_T$, with both changes occurring simultaneously at $t\approx 2.5$.

\begin{figure}
 \begin{center}
 \includegraphics[width=1\columnwidth]{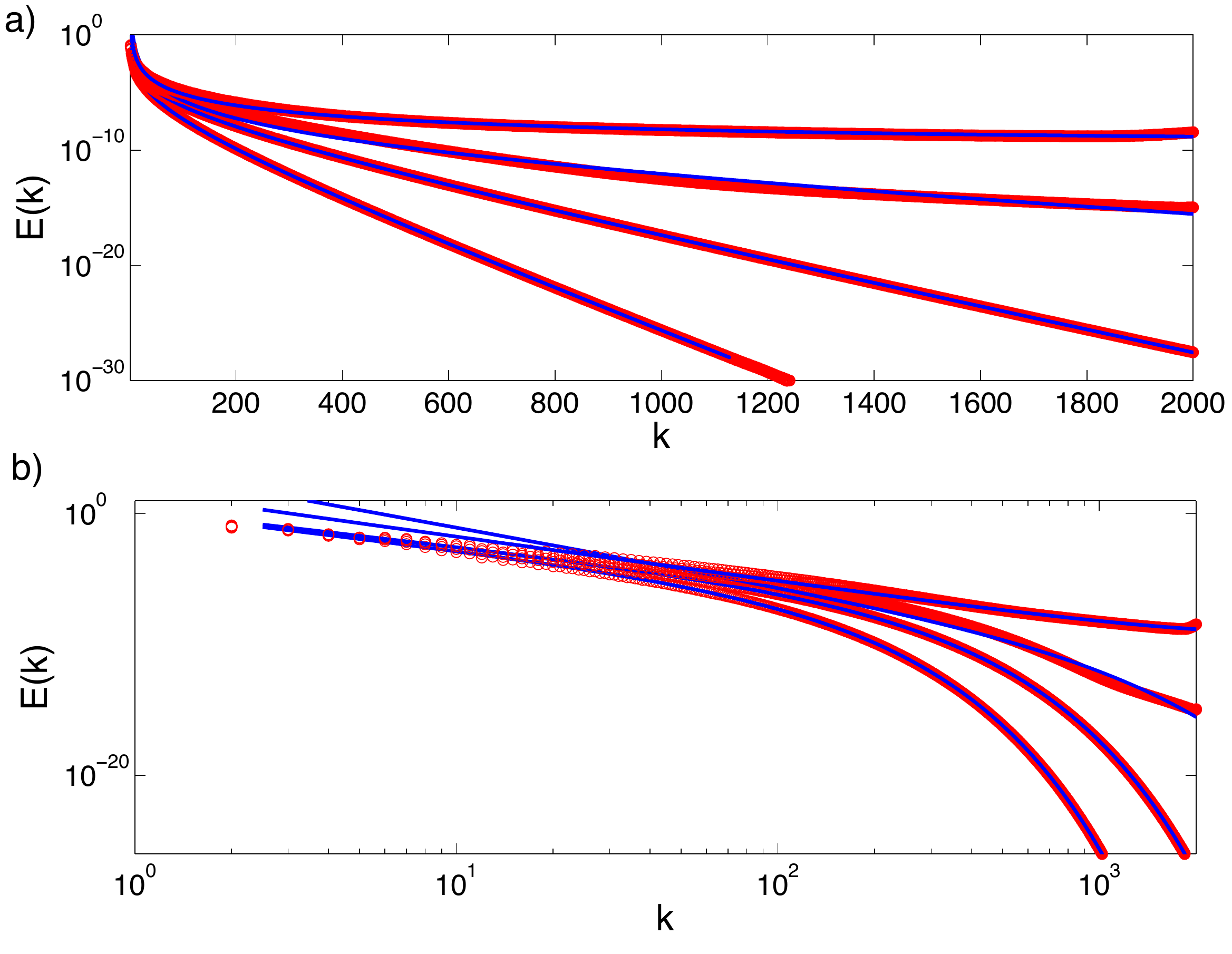}
 \caption{(color online) a) Fits (dark or blue lines) from $k=3$ to $k=k_{\rm max}$ using Eq.~\eqref{analytic} to the total energy spectra (light or red points), in lin-log scale, as a function of wavenumber and for different times: $t=1.975, \ 2.201, \ 2.425$ and $t=2.651$. Note the good quality of the fit at early times, and the poor quality at $t=2.425$. b) Same plot in log-log scale.
           }  \label{fig:fits}
 \end{center}
 \end{figure}

However, when comparing the fit using Eq.~(\ref{analytic}) to the actual spectrum in the simulations, one sees that errors are introduced as the spectrum is not always well represented by Eq.~(\ref{analytic}). This is associated with the fact that the simple form \eqref{analytic} needs to be true only in the $k \to \infty$ asymptotic. We now examine this point further.
In  simple flows such as the 1D Burgers solution corresponding to $\sin(x)$ initial data, or the purely hydrodynamic Taylor-Green vortex (see \cite{bustamante}), the energy spectrum of the flow can be globally well fitted with the simple form \eqref{analytic}, but this is not always the case. For instance, in the Kida-Pelz flow, oscillations were found and attributed to interferences of complex singularities, see \cite{cicho_05b}. In our simulation, the {insulating TG-MHD} total energy spectrum can be well fitted globally only up to $t=2.2$. After this time the energy spectrum displays a complicated behavior (see Fig. \ref{fig:fits}).

\begin{figure}
\begin{center}
 \includegraphics[width=1\columnwidth]{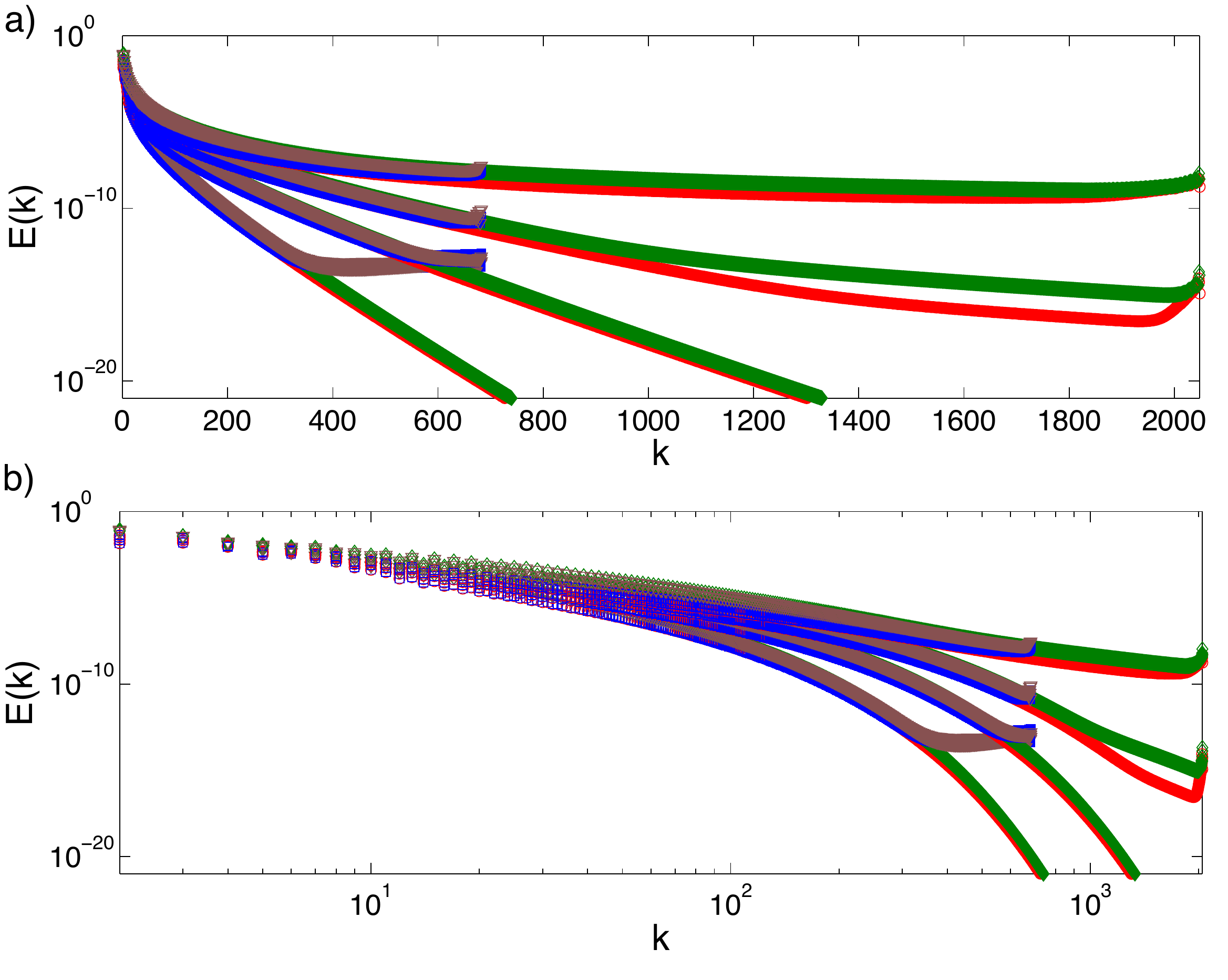}
 \caption{(color online)
Lin-log (a) and log-log (b) energy spectra as a function of wavenumber and for different times:
$t= 1.975, \ 2.201, \ 2.425$ and $2.651$, for the kinetic (light or red) and magnetic (dark or green) energy of the run at $6144^3$ resolution (the curves for this run go up to $k_{max}=2048$). Superimposed are the same for the ideal run in \cite{lee2}, at a resolution of $2048^3$ points in single precision (dark or blue for kinetic, light or brown for magnetic; the curves for this run go up to $k_{max}=682$). The same dominance of magnetic energy at small scales is observed in both runs; this implies extremely strong currents, compared to the vorticity, at small scales, as also observed when examining the temporal behavior of extrema (see Table \ref{tab:max}).
                }  \label{fig:2k6k}
 \end{center}
  \end{figure}

\begin{figure}
\begin{center}
\includegraphics[width=0.99\columnwidth ]{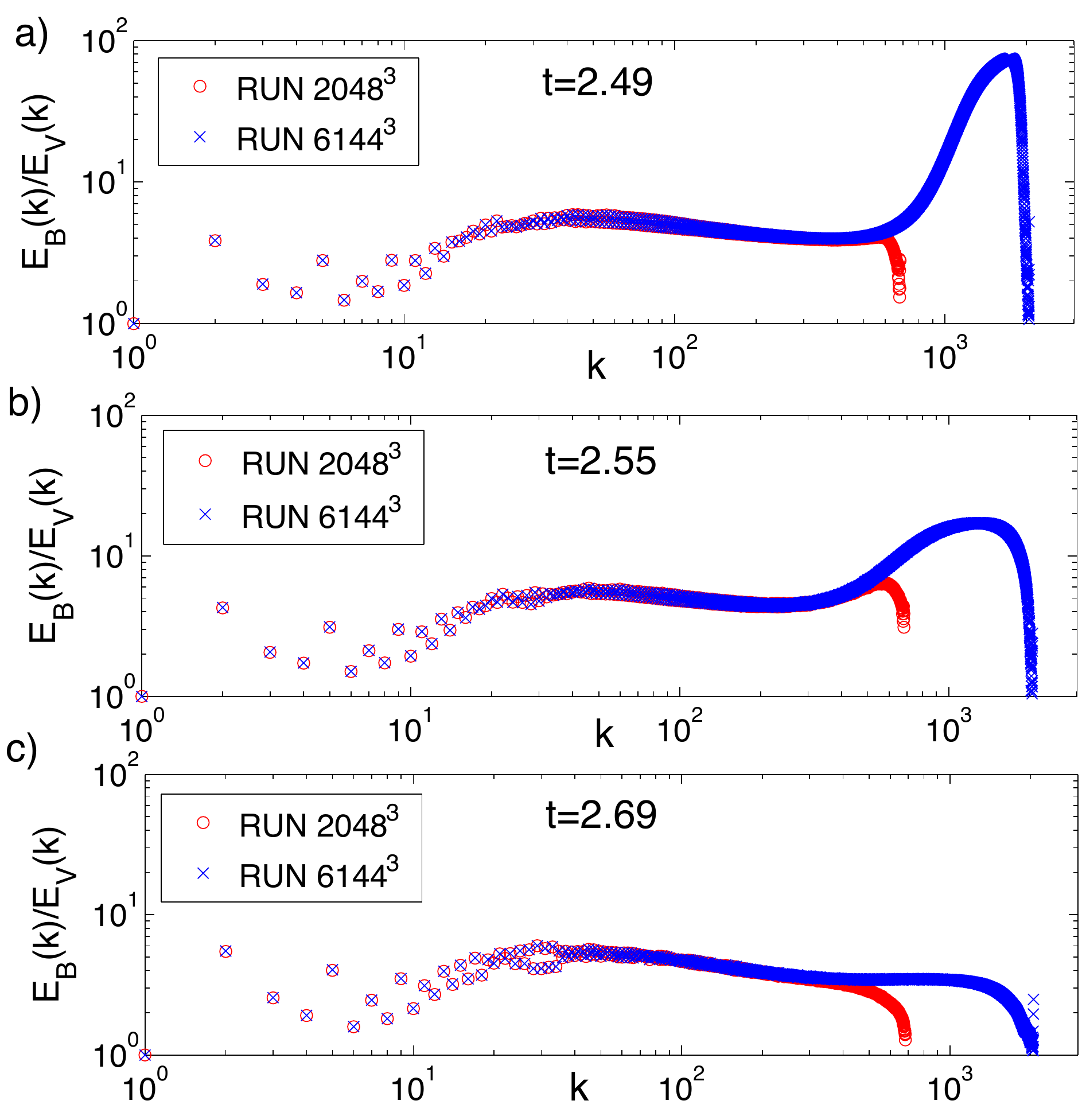}
\caption{(color online)
Ratio of magnetic to kinetic energy spectra for the I flow at three different times: t=2.49 (a), t=2.55 (b) and t=2.69 (c) for the run on a grid of $2048^3$ points (circles, red) \cite{lee2}, and on a grid of $6144^3$ (crosses, blue). The strong burst of excess magnetic energy at large wavenumbers subsides at later times.
 }  \label{fig:EMsEV} \end{center}
\end{figure}

To study if this is an effect associated with insufficient spatial resolution, in Fig.~\ref{fig:2k6k} we show the kinetic and magnetic energy spectra for the run performed on $6144^3$ points, as well as for a  run with the same initial conditions computed on a grid of $2048^3$ points {\it without} bootstrap re-gridding, and as analyzed in \cite{lee2}; we use both lin-log and log-log scales, for different times: $t= 1.975, \ 2.201, \ 2.425$ and $2.651.$
The implementation of the numerical procedure for the two runs in fact differs in several ways: (i) obviously, the resolution; (ii) single or double precision, the latter for the highest resolution; (iii) the truncation at high wavenumber (cubic for the latter, spherical for the former); and (iv) bootstrap regridding performed for the former, progressively in time. Yet, the two runs are seen to be equivalent.
As time progresses in these flows, the magnetic energy gains from its kinetic counterpart (remember that $E_V(t=0)=E_M(t=0)$), particularly so at high wave numbers, as is also shown in Fig.~\ref{fig:EMsEV}, which gives the variation with wavenumber of the ratio $E_M(k)/E_V(k)$ for three different times and for both the $2048^3$ and the $6144^3$ runs.
At $t\approx 2.48$, there is a surge of magnetic energy at small scales (large wavenumbers) compared to its kinetic counterpart, a surge which finally resolves itself at the final time of the computation. This behavior is likely linked to the evolution of structures in physical space (see \S \ref{ss:struct}).

\begin{figure}
 \begin{center}
 \includegraphics[width=0.99\columnwidth]{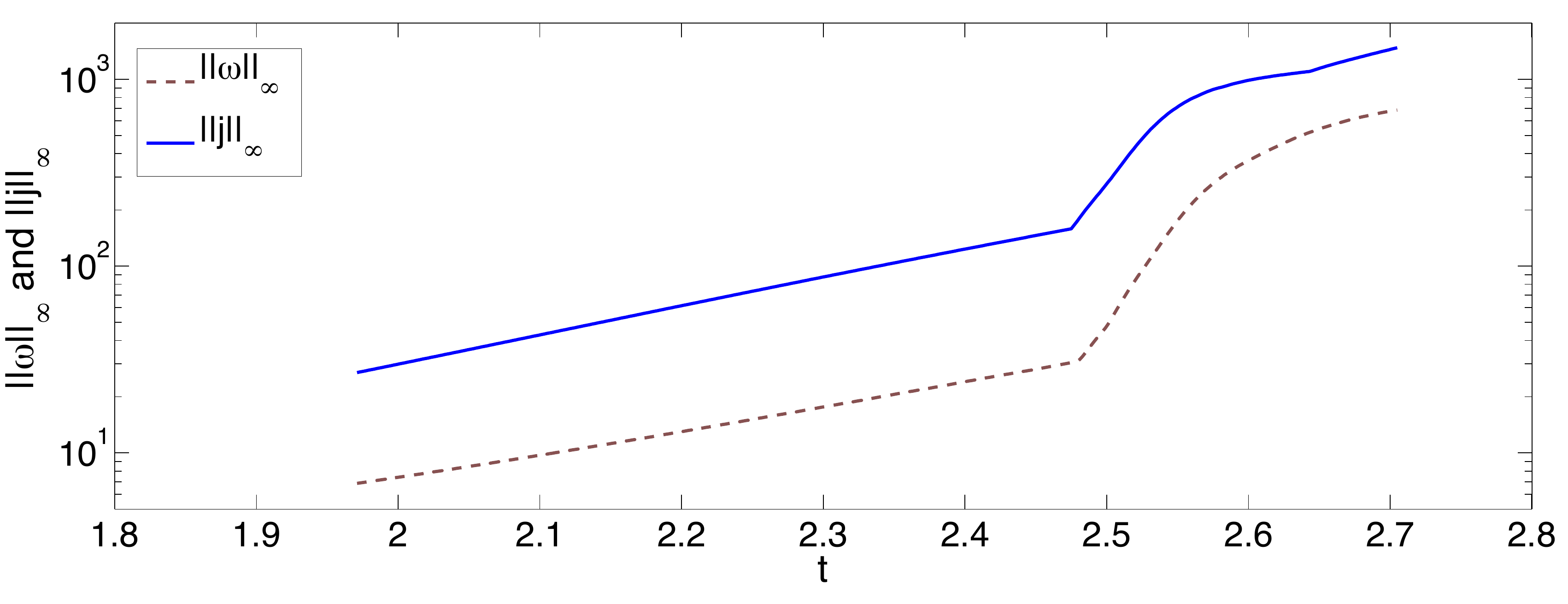}
 \caption{(color online)
 Maxima of vorticity (brown dashed line) and of current (blue solid line) as a function of time for the I flow at high resolution. Note the jumps of the slopes (see also Fig. \ref{fig:om_j_max} below) corresponding to the emergence of different leading structures.
               }  \label{fig:max}
  \end{center}
   \end{figure}

When investigating the temporal evolution of the vorticity and current maxima, as shown in Fig.~\ref{fig:max} (and also given in Table \ref{tab:max}), we observe that there is a sudden change in the slopes at $t\approx 2.5$, and again at $t\approx 2.65$, the latter clearly discernible in the current density. 
These changes are associated with a shift of the maximum from one structure to another one. The first phase of evolution, up to $t\approx 2.5$ is clearly exponential for both the maxima of current and the vorticity, followed by faster growth on new structures that appear at later times. However, because the strongest peaks in current and vorticity appear on a different structure at quite a late time in this run, when the grid resolution is almost reached, it is difficult to ascertain whether a singularity would happen or not in this flow if it were pursued to yet higher resolutions and thus longer times. In other words, due to the physical structures that develop in this flow, the traditional tests of singularity (BKM and logarithmic decrement) cannot be applied in the latest evolutionary phase because it is too short. From that point of view, computations on yet higher-resolution grids will be necessary. In the next subsection, we present a new analytical method that allows us to assess the plausibility of singularity scenarios.

\subsection{The link between the two known criteria for singularity} \label{ss:link}

It is known that several diagnostics for singularity can be used, and in fact that they are linked.
The first method is to follow the temporal evolution of the maximum of both vorticity and current and apply the BKM criterion given by Eq.~(\ref{BKM}) for fluids and generalized to the MHD case \cite{caflisch} (see \cite{klapper_98} for the two-dimensional case in MHD); for smoothness on the $[0, T]$ temporal interval, one must have convergence of the following integral:
\begin{equation}
\label{eq:BKM}
\int_0^T \left(||{\bf {\bomega}}(.,t)||_{\infty} + ||{\bf {j}}(.,t)||_{\infty}\right)\ dt  < \infty \ .
\end{equation}
Using the Els\"asser variables ${\bf z}^{\pm}={\bf v}\pm {\bf b}$ and defining the associated vorticities,
${\mathbf \bomega}^{\pm} = {\mathbf \bomega} {\pm} {\bf j}$, the above relation can also be written in characteristic form:
$$
\int_0^T \left(||{\mathbf {\bomega^+}}(.,t)||_{\infty} + ||{\mathbf {\bomega^-}}(.,t)||_{\infty}\right)\ dt  < \infty \ .
$$

Consider the formulation (\ref{eq:BKM}) for the BKM condition for regularity. If the numerical solution for the fields leads to a power-law behavior of the integrand, of the form $||{\bf {\bomega}}(.,t)||_{\infty} + ||{\bf {j}}(.,t)||_{\infty} \approx  C [T_\ast - t]^{-\beta}$, then the exponent $\beta$ must be greater than or equal to one in order to be consistent with the existence of a singularity at time $T_\ast.$

The second tool for singularity diagnostic is to follow the logarithmic decrement $\delta(t)$ of the fields mentioned above, in the context of the analyticity-strip method. In particular, one can look at the total energy spectrum (i.e., the spectrum of the sum of kinetic and magnetic energies) and calculate the decrement $\delta(t)$ for this spectrum. The logarithmic decrement $\delta(t)$ should go to zero in a finite time in order to be consistent with the existence of a singularity of the fields at time $T_\ast.$ In contrast, if $\delta(t)$ decays exponentially in time then there is no evidence for a finite-time singularity.
Finally, a third method consists of monitoring the evolution of the total production of small scales, through the enstrophy (integrated square vorticity) and the integrated square current.

It may appear a bit odd to have different criteria to determine the evolution or not towards a singularity, but this is not redundant; quite the contrary.
The link, at the level of heuristics, between the enstrophy divergence and that of vorticity was shown in \cite{kerr_05, kerr_08}. More recently, a rigorous proof that bridges the two other criteria for singularity (BKM theorem and analyticity strip method) was shown in \cite{bustamante} along with an application to a numerical simulation of a 3D Euler fluid. The advantage of this bridge is that it leads to a new criterion when monitoring of the temporal evolution of small scales, giving an inequality between the power-law index of the energy spectrum and the temporal index of evolution for the logarithmic decrement, provided they can be assessed reliably.

To this end, one needs to use known inequalities. For our purposes, we recall the result in \cite{bustamante} that links the maximum vorticity modulus with the 3D Euler energy spectrum:
\be
\label{eq:link}
||{\bf {\bomega}}(.,t)||_{\infty} \le c\,\sum_{k = 1}^{\infty} k^2 \sqrt{E(k,t)} \ , \ \ \forall t \in [0,T) \ ,
\ee
where $c$ is a constant of $\mathcal{O}(1).$

The key concept in this new bridge is a hypothetical bound for the energy spectrum of the form
\be
\label{eq:bound}
E(k,t) \leq M k^{-n_0(t)} e^{-2 k \delta_0(t)}\,, \quad \forall \, t \in [0,T)\,, \quad \forall \,k \in \mathbb{N},
\ee
for certain positive functions $n_0(t)$ and $\delta_0(t),$ and some positive constant $M$. The functions $n_0, \delta_0$ are closely related to the analyticity-strip fit parameters $n_X$, $\delta_X$ considered above, but they are not the same. In fact, the above hypothetical bound is global (in $k$-space), whereas as already mentioned the logarithmic decrement $\delta_X(t)$ gives information on the asymptotic (large-$k$) behavior of the energy spectrum.

It was demonstrated in \cite{bustamante} that combining this hypothetical bound with the rigorous inequality (\ref{eq:link}) leads to a relation between the BKM theorem and the analyticity-strip method. To simplify matters, one considers the consequences of the following finite-time singularity scenario: suppose for simplicity that the exponent $n_0$ in the hypothetical bound (\ref{eq:bound}) remains constant as $t$ approaches the singularity time $T_*$, and that $\delta_0(t) \propto (T_{*}-t)^{\gamma}$, where $\gamma >0$. Then the following necessary condition is found:
$$
\gamma  \ge \frac{2}{6-n_0}, \
$$
in order that the blow-up be consistent with the BKM theorem. The formal argument is given in \cite{bustamante} and is immediately generalizable to MHD. The result for MHD is as follows:
\be
\label{eq:linkMHD}
||{\bf {\bomega}}(.,t)||_{\infty} + ||{\bf {j}}(.,t)||_{\infty} \le  c\,\sum_{k = 1}^{\infty} k^2 \sqrt{2 E_{\scriptscriptstyle{\mathrm{T}}}(k,t)} \ , \ \ \forall t \in [0,T) \ ,
\ee
where now $E_{\scriptscriptstyle{\mathrm{T}}}(k,t)$ represents the total energy spectrum, i.e., the sum of kinetic and magnetic energy spectra. The corresponding hypothesis for energy bound (\ref{eq:bound}) is unchanged and similarly the hypothesis of blow-up for $\delta_{\scriptscriptstyle{\mathrm{T}}}(t).$ The result is again a necessary condition, of the form
\be
\label{eq:gamma_bound}
\gamma_{\scriptscriptstyle{\mathrm{T}}}  \ge \frac{2}{6-n_{\scriptscriptstyle{\mathrm{T}}}}.
\ee
Note that, since in the Euler case, the observed $n_0$ appears to be (at least for some initial conditions) larger than the exponent $n_{\scriptscriptstyle{\mathrm{T}}}$ in the MHD case, one sees that the eventual realization of a singularity in MHD might be a different process than for the Euler equation. This is not necessarily surprising for at least three reasons: (i) MHD is thought to be smoother than hydrodynamics, insofar as Alfv\'en waves may slow down the dynamics of propagation to small scales, leading possibly to a different energy spectrum, the so-called Iroshnikov-Kraichnan law;  (ii) the Onsager principle concerning energy dissipation can likely be replaced in MHD by magnetic helicity conservation, following the so-called Taylor conjecture \cite{caflisch}, thereby changing the dimensionality of the system; and (iii) the degree of smoothness required to ensure total energy conservation (technically, the index of the Besov space needed)  for the velocity and the magnetic field may differ in a way that is compatible with the Iroshnikov and Kraichnan spectra \cite{caflisch}.
In particular, with  $n_0\approx 4$ for Euler, one obtains $\gamma \geq 1$ whereas for $n_{\scriptscriptstyle{\mathrm{T}}}\approx 3$ in ideal MHD (see \cite{lee_ideal}), one has $\gamma_{\scriptscriptstyle{\mathrm{T}}} \geq 2/3$: the decay of the logarithmic decrement would be slower in MHD, as expected because of the slowing-down of the dynamics by (Alfv\'en) waves.

\subsubsection{Analysis of the total energy spectrum}

At time $t\approx 2.33$, we observe a change in the behavior of the total energy spectrum, probably due to the imminent, accelerated collision between two current sheets (confirmed by inspection of the structures in real space), and the corresponding fast generation of a second length scale, related to the distance between the two sheets. The original length scale of the problem, interpreted as the decreasing width of the current sheets, decreases slower than this new length scale so eventually the two length scales become comparable. It is known that when two or more sharp physical structures of similar length scales are present, the traditional fit (\ref{analytic}) of the energy spectrum fails. For example, in the Kida-Pelz 3D Euler flow, the departure of the measured energy spectrum from the traditional form (\ref{analytic}) was modeled with good accuracy by attributing it to interferences of two complex singularities \emph{situated at equal distances from the real axis} \cite{cicho_05b}. However, the extra complexity (spatial and temporal) of the MHD flow under current study makes it difficult for us to find a good model for this new behavior. This imposes a practical limitation on the analyticity-strip method as a means for finding a good estimate of the actual logarithmic decrement $\delta_{\scriptscriptstyle{\mathrm{T}}}(t)$ of the spectrum (where ``actual'' is used in contrast to the measured one). In fact, depending on the fit interval we get vastly different estimates for the width $\delta_{\scriptscriptstyle{\mathrm{T}}}(t)$ for times $t>2.33,$ so our knowledge of the  width $\delta_{\scriptscriptstyle{\mathrm{T}}}(t)$ as in the large-$k$ asymptotic expansion $\ln E_{\scriptscriptstyle{\mathrm{T}}}(k,t) \sim -2\,k\,\delta_{\scriptscriptstyle{\mathrm{T}}}(t)$ has significant errors that grow in time.

\begin{figure}
\begin{center}
 \includegraphics[width=1\columnwidth]{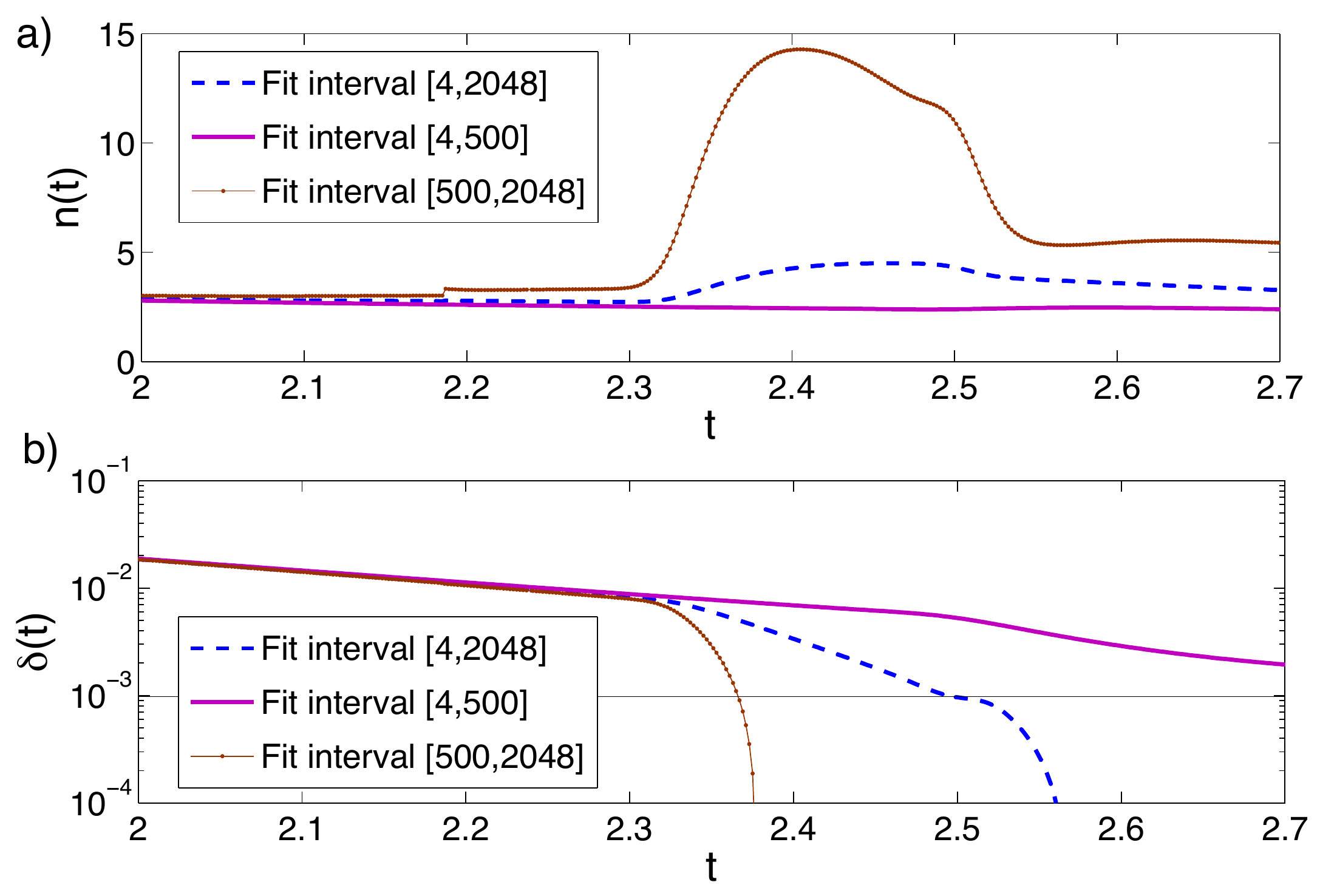}
 \caption{(color online)
Running estimates of the analyticity-strip method exponents $n_{\scriptscriptstyle{\mathrm{T}}}(t)$ (a) and $\delta_{\scriptscriptstyle{\mathrm{T}}}(t)$ (b) for the total energy spectrum. In dashed (blue) line for a fit interval $[4,2048]$; in thick solid (magenta) line for a fit interval $[4,500]$; in thin with dots (brown) line for a fit interval $[500,2048].$ In (b), the horizontal line represents the reliability threshold $\delta_{\scriptscriptstyle{\mathrm{T}}}  k_{\max} = 2.$
 }  \label{fig:fig3c}
  \end{center}
  \end{figure}

In conclusion, we cannot tell by using the analyticity-strip method alone whether there is a finite-time singularity in the MHD flow under study at times  $t>2.33.$ Of course, we know from continuity arguments that the width $\delta(t)$ should remain non-zero at least for a short time after $t=2.33.$ But that is all we know, so there are two possible scenarios:\\
\begin{description}
\item[Scenario 1:] There is no finite-time singularity up to time $t=2.7,$ so the simulation is well resolved, perhaps marginally. The implications of Scenario 1 will be exploited in Section \ref{ss:struct}.\\
\item[Scenario 2:] There is a finite-time singularity at a time between $t=2.33$ and $t=2.7,$ but this cannot be assessed using the analyticity-strip method alone.\\
\end{description}
Let us consider the implications of Scenario 2. Although we do not know the logarithmic decrement $\delta_{\scriptscriptstyle{\mathrm{T}}}(t),$ we can still have an estimate for the positive exponent $n_{\scriptscriptstyle{\mathrm{T}}}(t)$ appearing in the bound (\ref{eq:bound}) for the total energy spectrum. In fact, what is needed in inequality (\ref{eq:gamma_bound}) is a lower bound for $n_{\scriptscriptstyle{\mathrm{T}}}(t),$ rather than $n_{\scriptscriptstyle{\mathrm{T}}}(t)$ itself. This lower bound can be estimated by looking at the low wavenumber fits of the total energy spectrum (from $k=4$ to $k=500$), as shown in Fig.~\ref{fig:fig3c}. Running estimates for $n_{\scriptscriptstyle{\mathrm{T}}}(t)$ obtained in this way turn out to be consistently smaller than the estimates obtained by using fit intervals including larger values of $k$. The result for the lower bound is $n_{-} = 2.385.$  With this number, the inequality (\ref{eq:gamma_bound}) gives a bound for the unknown exponent $\gamma_{\scriptscriptstyle{\mathrm{T}}}$ in $\delta_{\scriptscriptstyle{\mathrm{T}}}(t) \propto (T_*-t)^{\gamma_{\scriptscriptstyle{\mathrm{T}}}}:$
\begin{equation}
\label{eq:gbound}
\gamma_{\scriptscriptstyle{\mathrm{T}}} \geq \frac{2}{6-n_{-}} \approx  0.553 \, ,
\end{equation}
so even though we do not know whether the logarithmic decrement is going to zero or not in a finite time, we have been able to estimate how fast it should go to zero in the hypothetical case of a finite-time singularity.\\

\subsubsection{Analysis of the sum of supremum norms of vorticity and current}

 \begin{figure}
  \begin{center}
 \includegraphics[width=1\columnwidth]{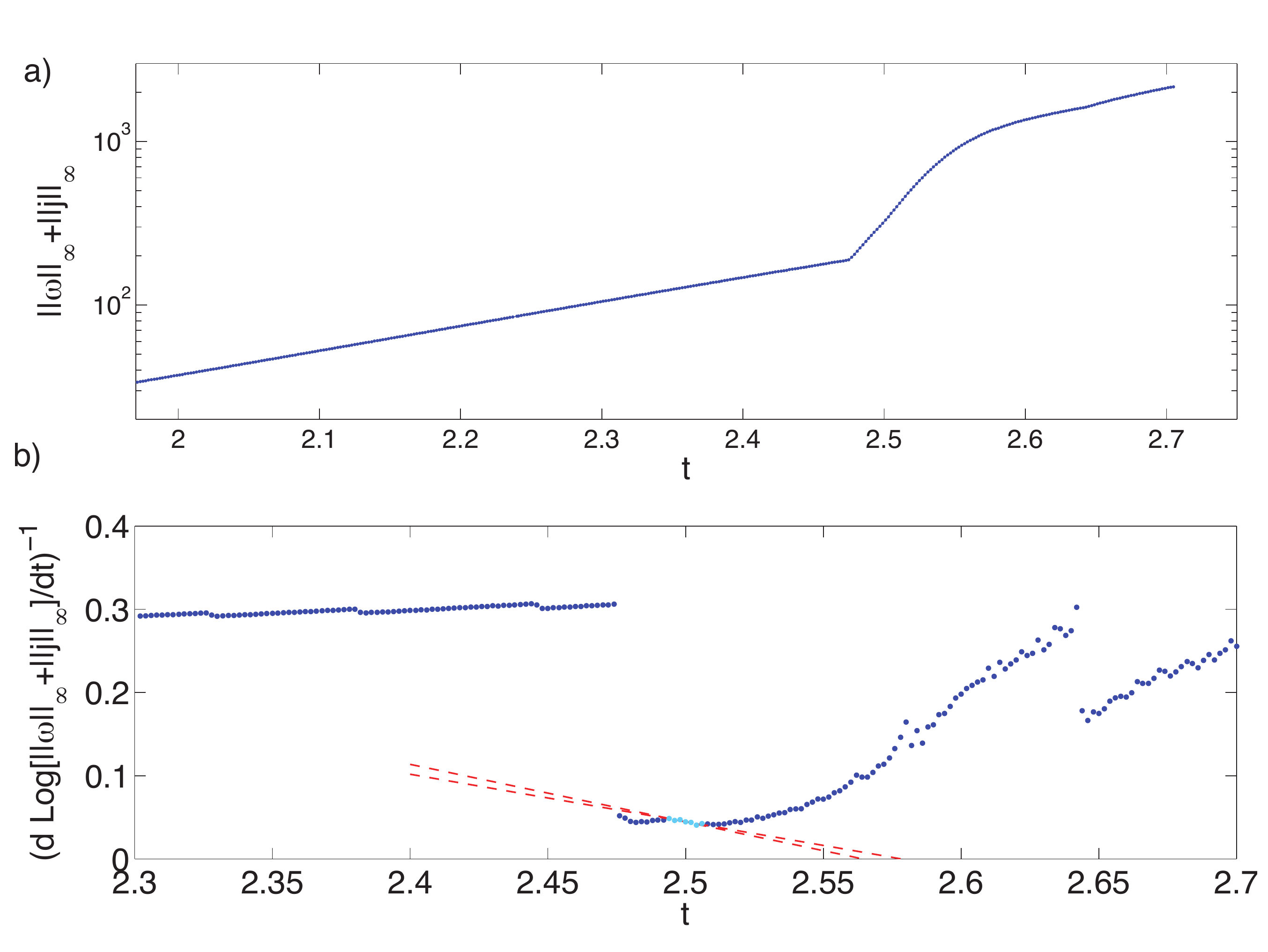}
 \caption{(color online)
a) Sum of maxima of vorticity and current as a function of time. There is a clear jump at $t=2.476$ corresponding to an emergent near-singular structure taking over a previous one. b) Multiplicative inverse of the logarithmic derivative of the previous curve. If this has a negative slope, it is  an indication for a possible finite-time singularity.
}  \label{fig:om_j_max}
 \end{center}
\end{figure}

To further comment on the feasibility of Scenario 2, let us consider the method of running estimates for singularity of fast-growing quantities  introduced in \cite{kerr_08}. We apply this method to the growth of the BKM field $||{\bf {\bomega}}(.,t)||_{\infty} + ||{\bf {j}}(.,t)||_{\infty}$ with the ansatz $||{\bf {\bomega}}(.,t)||_{\infty} + ||{\bf {j}}(.,t)||_{\infty} \approx  C [T_\ast - t]^{-\beta}.$ The method gives running estimates of the exponent $\beta$ and of the singular time $T_*$.
In Fig.~\ref{fig:om_j_max}(a) we observe that there is a jump at $t=2.476$ in the growth rate of the BKM quantity; however, this is not due to a dynamical effect. It is rather due to an independently emergent physical structure that is more singular than the previous one.
Figure \ref{fig:om_j_max}(b) shows the multiplicative inverse of the logarithmic derivative of the BKM quantity. If this curve has negative slope, then the intersection of the slope with the $t$-axis gives a running estimate of the potential singularity time. We see two instances of negative slope. We discard the instance at about $t=2.476$ because this is due to the transient emergence of the new structure. However, near $t=2.5$ we observe more robust evidence of potential singularity, although the data is quite noisy and thus the tangent is oscillating too much, so we cannot have precise estimates of the singularity time and the exponent $\beta$. Naked-eye prediction of singularity time, obtained by finding the intersection of the smoothed tangent at $t=2.5$ with the $t$-axis, would give $T_* \approx 2.56$--$2.58$ and $\beta \approx 1.44$--$1.75.$ These values for the estimates of $T_*$ and $\beta$ were obtained by estimating two tangents in Fig.~\ref{fig:om_j_max} (b), each tangent being defined as the linear interpolation of six contiguous data points taken out of the seven data points highlighted in the figure.

It is interesting that near $t=2.5$ the estimated logarithmic decrement $\delta_{\scriptscriptstyle{\mathrm{T}}}(t)$ (using the full fit range $[4,2048]$) indeed has a change in behavior, first a deceleration and then an acceleration, although this occurs near the reliability threshold--see Fig.~\ref{fig:fig3c}(b). A computation of the running estimate of decay exponent $\gamma_{\scriptscriptstyle{\mathrm{T}}}$ as in $\delta_{\scriptscriptstyle{\mathrm{T}}}(t) \propto (T_*-t)^{\gamma_{\scriptscriptstyle{\mathrm{T}}}}$ gives $\gamma_{\scriptscriptstyle{\mathrm{T}}} = 0.94$ at $t=2.501$ but only that data point agrees with the rigorous bound in (\ref{eq:gbound}), $\gamma_{\scriptscriptstyle{\mathrm{T}}} \geq 0.553.$ At slightly later times, the estimated value of $\gamma_{\scriptscriptstyle{\mathrm{T}}}$ becomes 10 times smaller, thus violating the rigorous inequality. The corresponding predicted singular time, using this method, gives a running estimate $T_*\approx 2.516$--$2.522.$

To summarize, Scenario 2 is plausible but some of its aspects occur in the limit of the reliability threshold. This point is aggravated by the fact that the sampling of current and vorticity maxima at the grid points induces spurious oscillations in the data (a way to suppress these oscillations is discussed in Sec.~\ref{s:conclu}). Therefore, no robust conclusion can be drawn at the moment. A future higher-resolution numerical simulation should shed more light on the feasibility of Scenario 2.\\

\begin{table} \begin{tabular}{| c  || c | c | c | c || c | c |c|c|}
\hline
  Time \  & \ $10^{-3}J_{max}$ &$ i_J$\  & $j_J$ \ & $k_J$\  \ & \  $10^{-3}\omega_{max}$ & $i_{\omega}$\   & $j_{\omega}$\  & $k_{\omega}$ \ \\
  \hline
  2.40  & $15.2$    & 1  & 1            & 1231   & $0.58$  & 336 & 7 & 1275 \\
  2.45  & $21.3$    & 1  & 1            & 1266   & $0.79$  & 342 & 6 & 1303 \\
  \hline
  2.50  & $75.9$    & 1  & 72            & 1522   & $2.3$  & 40 & 93 & 1518 \\
  2.55  & $508$    & 1  & 95            & 1520   & $30.7$  & 40 & 107 & 1519 \\
  2.60  & $972$    & 1  & 153            & 1508   & $135$  & 70 & 140 & 1519 \\
  \hline
  2.65  & $1312$  & 62  & 62            & 1537   & $294$  & 128 & 172 & 1525 \\
  2.70  & $20886$ & 81  & 81            & 1537   & $456$  & 197 & 226 & 1528 \\

    \hline  \end{tabular}   \caption{
     Time, maxima of current, and their $(i_J, j_J, k_J)$ location in the fundamental $[0,\pi/2]$ box in grid units, as well as maxima of vorticity and their location, for the high-resolution I flow on a grid of $6144^3$ points. The indices $(i, j, k)$ refer to the grid points in $(x,y,z)$ where the maxima take place. Note the sudden jumps in the position of the maxima; the first jump in coordinates and in values of maxima occur for $t\approx 2.48$, and the second one at $t\approx 2.62$.
} \label{tab:max}   \end{table}

\subsection{Structures in physical space} \label{ss:struct}

Visualization plays an important role in the discovery process, and many of the arguments considered above used information from the evolution of the structures in physical space, based on previous runs \cite{lee_ideal} and the present high-resolution computation. In order to visualize the velocity and magnetic field and their gradients, one needs to reconstruct the three-dimensional data using the four-fold symmetries of the TG-MHD configuration, a daunting task at such resolutions. In that context, note that the VAPOR visualization system developed at NCAR \cite{vapor, mininni_08} allows one to analyze the data using wavelet compression in order to explore rapidly at coarser resolutions, and then to increase the resolution as needed where needed.

The acceleration in the formation of small scales was first identified in \cite{lee_ideal} with the collision of two current sheets leading to a quasi-rotational discontinuity. The present computations at higher resolution confirm these results and allow us to go further in time. We have given in Table \ref{tab:max} the values close to the end of the computation of the maximum of the vorticity and of the current as well as their location in the fundamental $[0,\pi/2]$ computational box. Concentrating on the current, which is known in two dimensions to have a simpler geometric structure (a dipole instead of a quadrupole for the vorticity), we observe two jumps, near $t_1=2.48$ and $t_2=2.62$, both in the value of the current maximum and in its location.

The collision of two current sheets leading to a  quasi-rotational discontinuity was clearly observed in \cite{lee_ideal} at a resolution of $2048^3$ points; this phenomenon is confirmed in the present computation with three times the linear resolution and thus any numerical effect can be ruled out for it. The second acceleration in the development of small scales, which occurs at a later time, seems to be related to the near co-location of these two sheets and this is now what we examine by considering structures in physical space. We also note that such current sheets are known to roll-up at sufficiently high resolution in the dissipative case, and similar rolled-up structures have been observed in the Solar Wind in a much more complex physical environment \cite{hasegawa_04, phan_06}.  But they are only a recent finding in DNS of MHD turbulence on grids of $1536^3$ points using the GHOST code \cite{mininni_06a, mininni_06b, mininni_09} and $2048^3$ (equivalent) points using TYGRS \cite{gafd}.

The maximum of the current first moves along the vertical axis (index $k$), then at $t=2.5$ it has moved in the y direction (index $j$): it is traveling  along the lower sheet as the two sheets seem to join with each other, to follow the curvature. Then finally, around $t=2.65$, the current density maximum now moves along the diagonal in the horizontal plane.

\begin{figure*}
\begin{center}
\includegraphics[width=0.65\columnwidth]{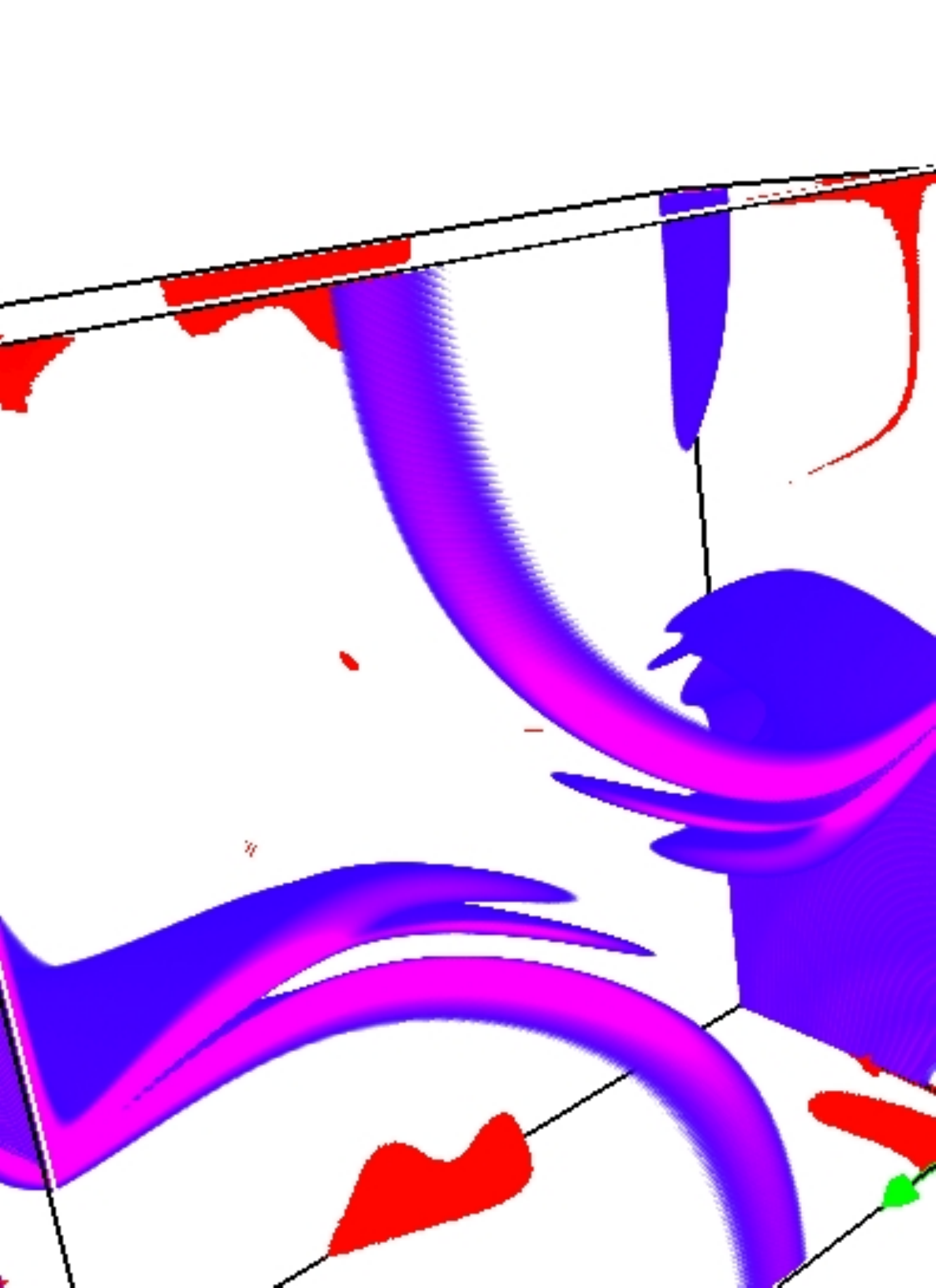}
\includegraphics[width=0.65\columnwidth]{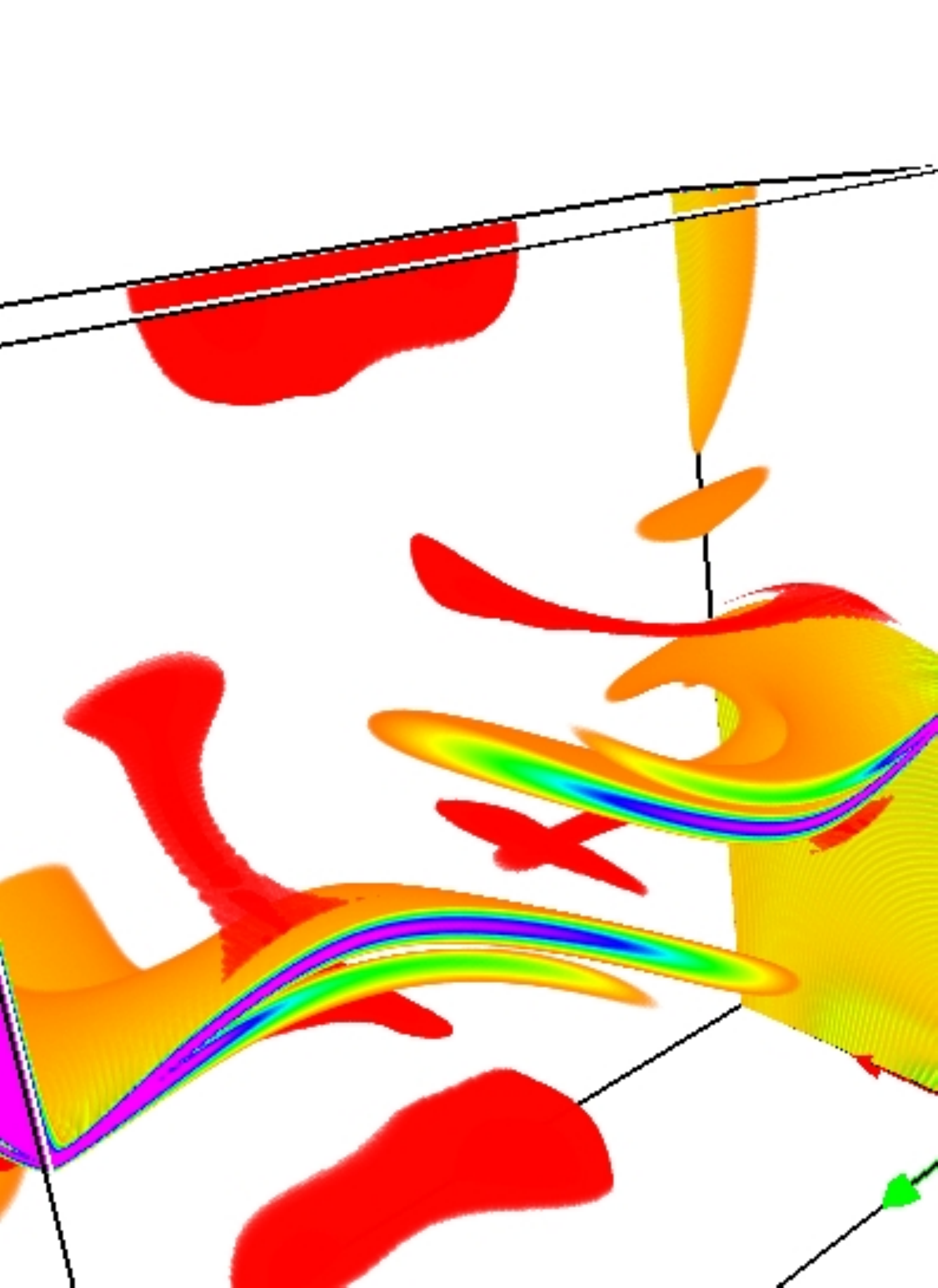}
 \includegraphics[width=0.65\columnwidth]{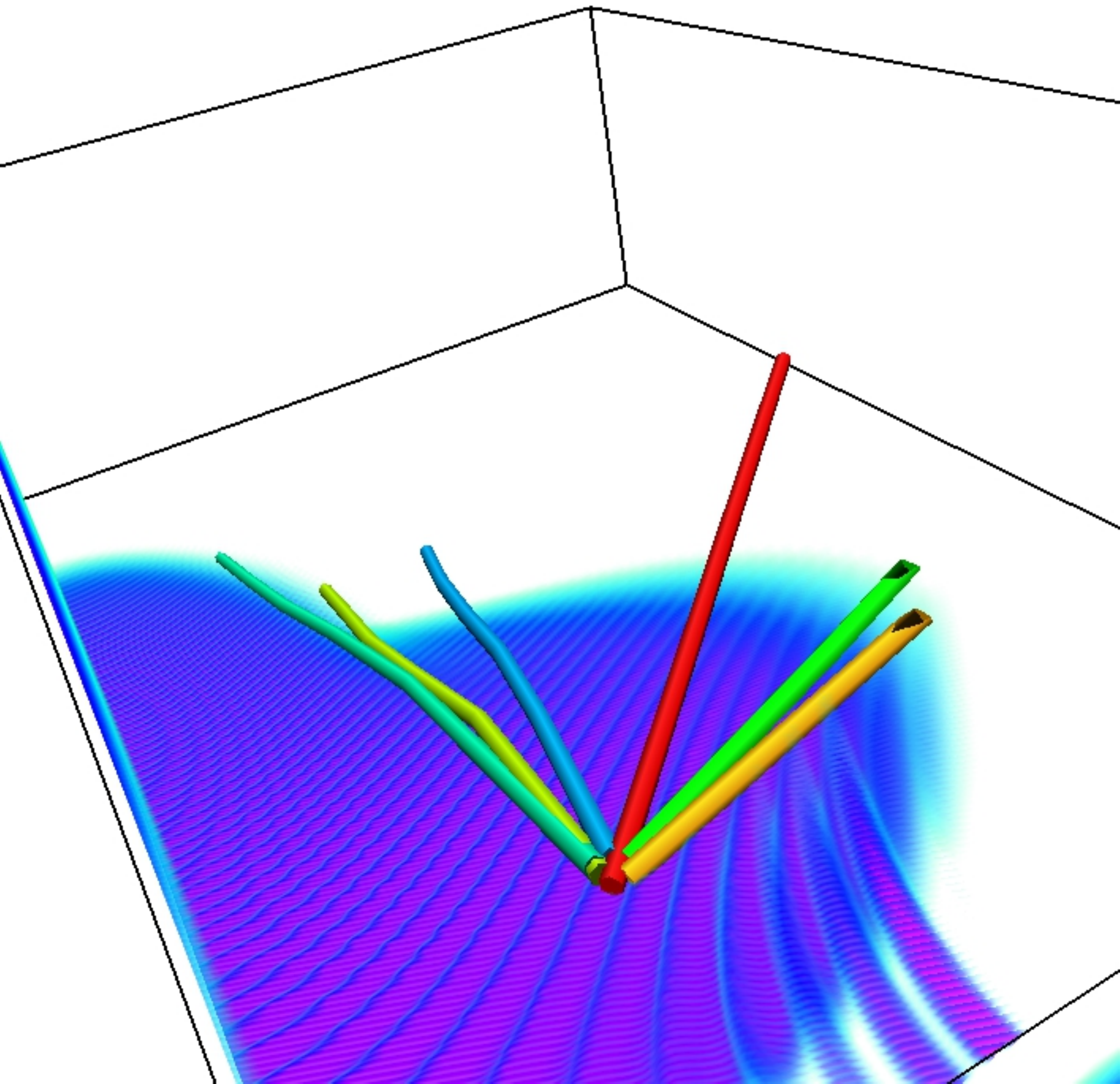}
 \caption{(color online)
Perspective volume rendering using the VAPOR software \cite{vapor, mininni_08} of the vorticity (left) and of the current density (middle) at t=2.54. Note the occurrence of a double layer structure due to the collision and subsequent joining of two sheets. At a later time (t=2.65; right), the magnetic field lines taken on these two colliding sheets all go to the same location, which coincides with the maximum of the current (coordinates given in the Table), implying sharp localized bending (and possibly torsion) of magnetic field lines in the vicinity of that maximum.
 }  \label{fig:pvr} \end{center}
  \end{figure*}

 \begin{figure} \begin{center}
 \includegraphics[width=0.45\columnwidth]{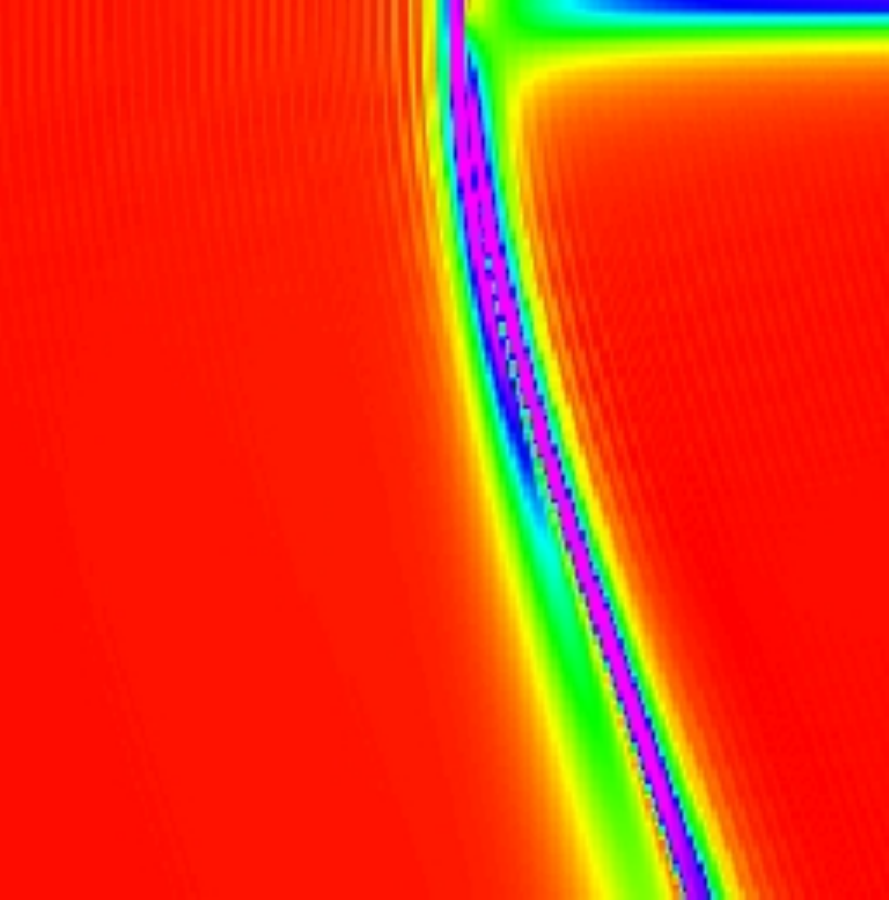}
 \includegraphics[width=0.45\columnwidth ]{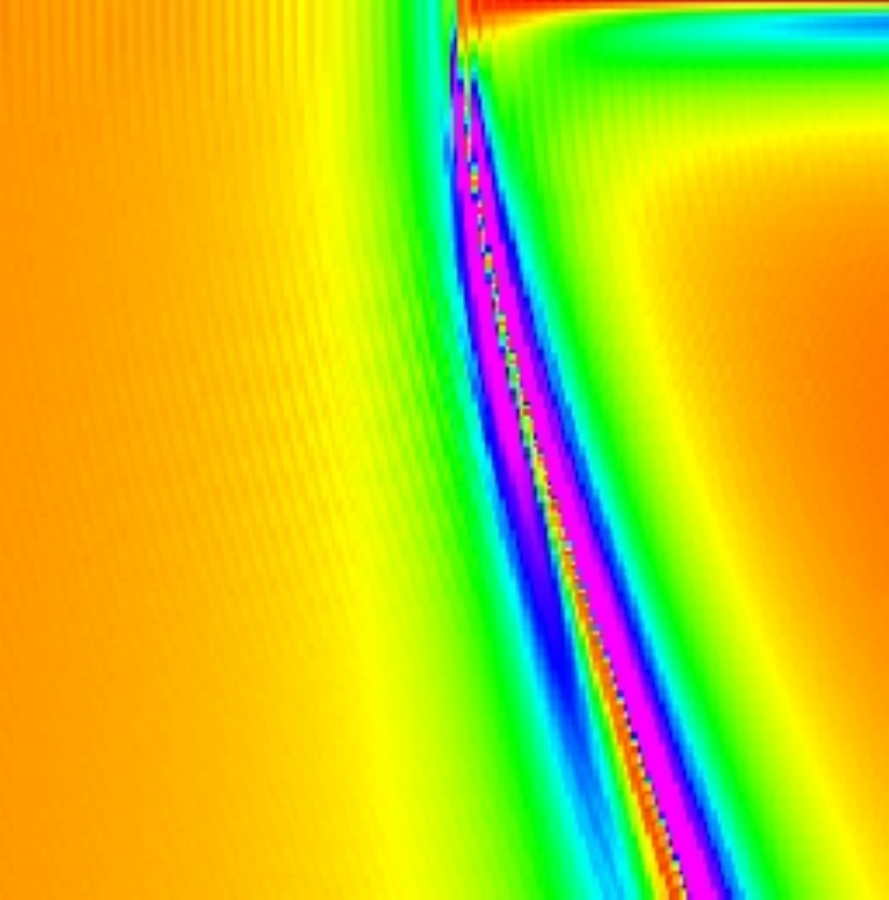}
  \caption{(color online)
Alternative views of the same structures at the same time as in Fig. \ref{fig:pvr}: two-dimensional cuts of vorticity (left)  and of current (right)  in the region of the current sheet collision, displayed down to the grid-resolution.
 }  \label{fig:pvr3} \end{center}  \end{figure}

A rendering of the current is given in Fig. \ref{fig:pvr} at t=2.54 (left and middle for the vorticity and the current).
It is the merging of two current sheets that causes the maximum to go radially (in a cylindrical sense) from the corner of the fundamental box ($[i,j,k]=[1,1,1537]$) along a polar angle of $\pi/2$ on the top plane. The two sheets (seen at both ends of the box because of symmetries) are clearly almost touching each other (a zoom indicates that they are two to three grid-points apart). Finally, on the right is given a suite of six magnetic field lines at the latest time of the computation that appear to all converge to one point, indicative of a potential singularity, that point being the location of the current maximum at that time. The two current sheets are barely visible (purple, and blue below). Alternative views are given in Fig. \ref{fig:pvr3}, with in particular two-dimensional cuts at the highest (grid) resolution, indicating that the two current and vorticity sheets are still individually resolved.

 Such  features correspond to a strong bending of magnetic field lines in the vicinity of the current and vorticity maxima, implying strong directional variations.
It may also imply magnetic field line stretching in this strong curvature region, a stretching that would be consistent with the sudden increase in magnetic energy (relative to its kinetic counterpart), as observed clearly (see Fig. \ref{fig:EMsEV}).
This is also reminiscent of the necessity, in the Euler case, of a blow-up of both the magnitude of the small-scale field but also of the curvature of its field lines, as shown in  \cite{constantin1994, constantin2008}, for a singularity to occur.

\subsection{The case of other Taylor-Green configurations in ideal MHD} \label{ss:IAC}

\begin{figure}
\begin{center}
 \includegraphics[width=0.99\columnwidth]{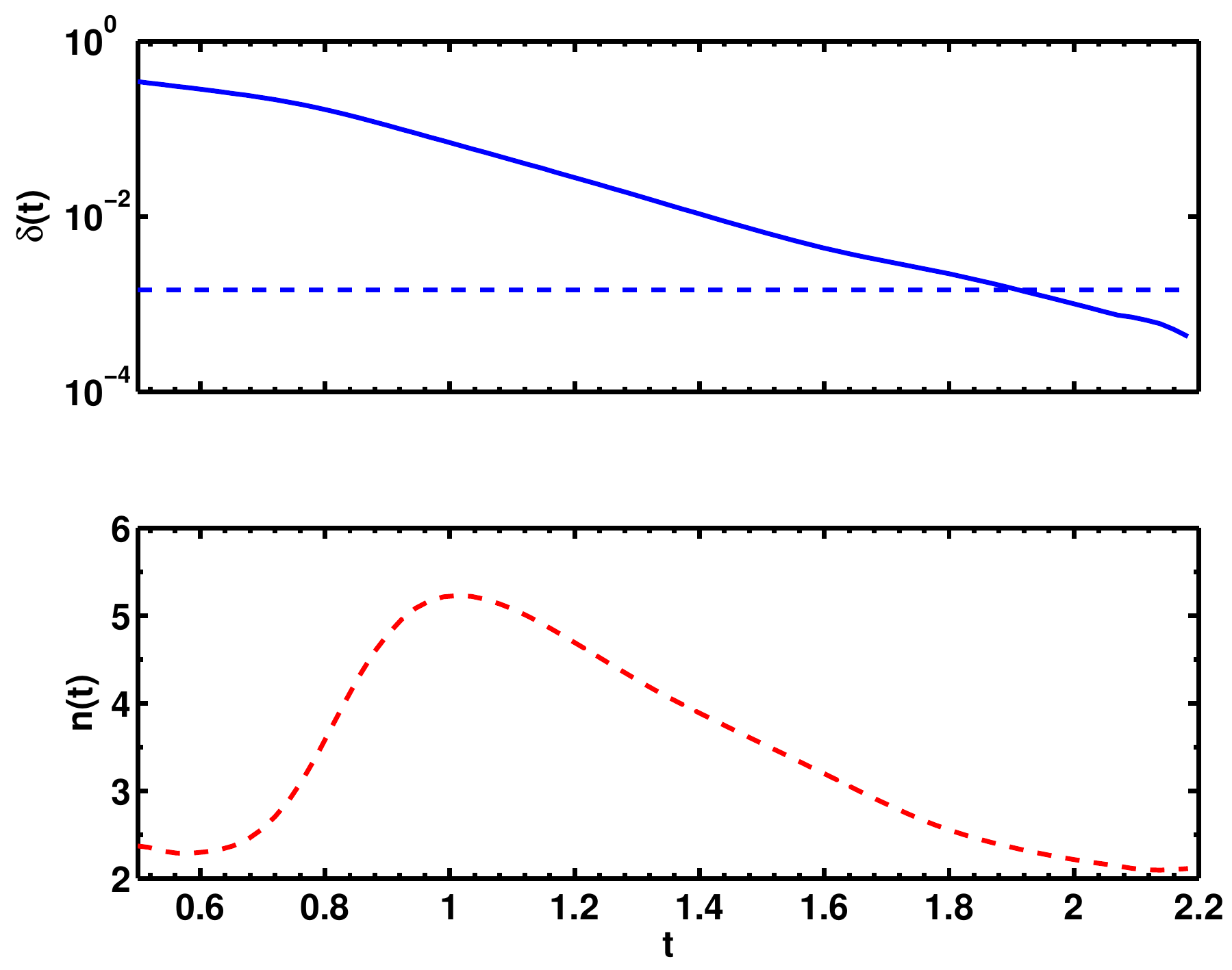}
    \includegraphics[width=0.99\columnwidth ]{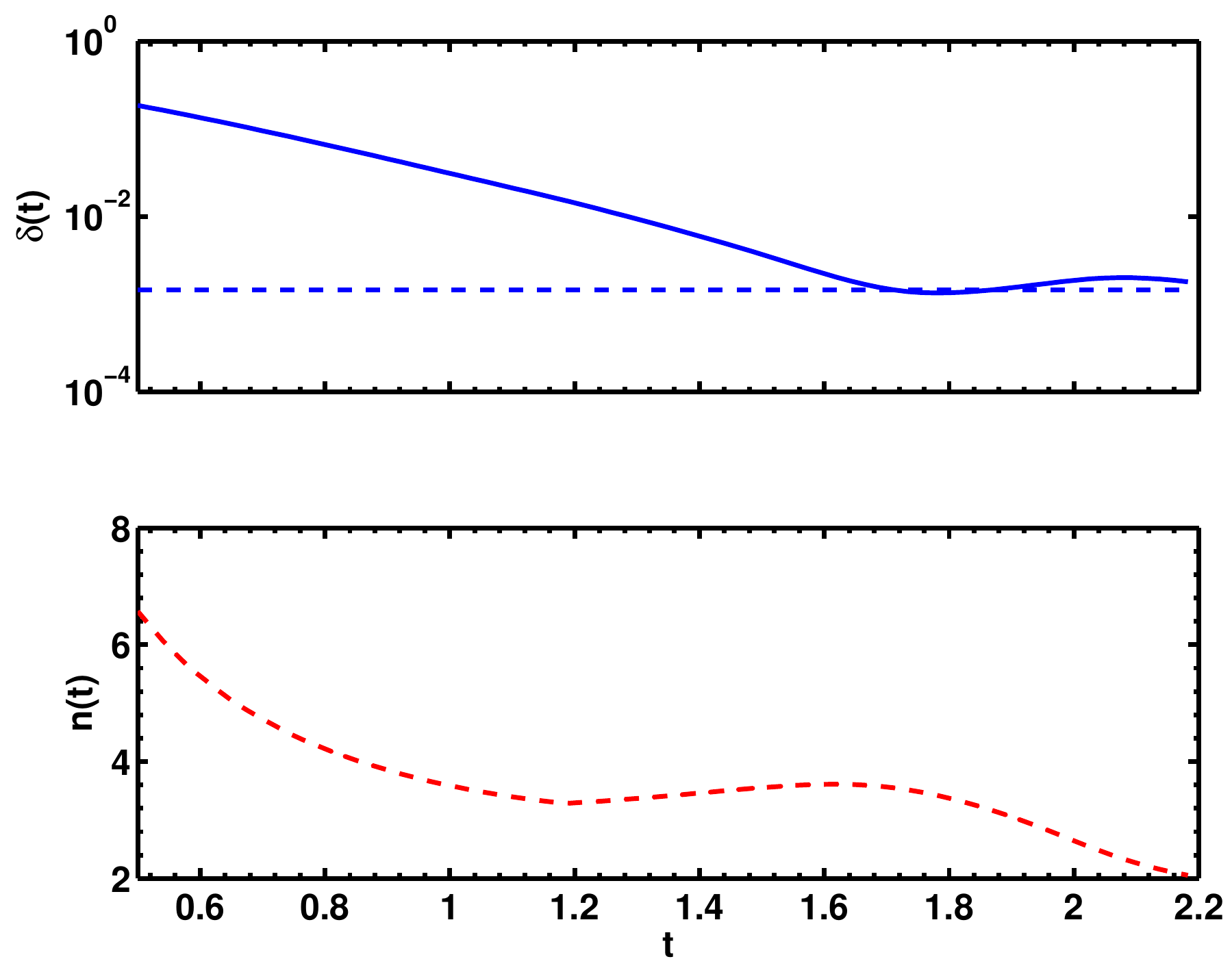}
  \caption{(color online) Analysis of the fit to the total energy spectra for the ``A'' flow (two upper panels) and ``C'' flow (two lower panels) on grids of up to $4096^3$ points; note that early times are not shown.
{\it First and third panels:} Logarithmic decrement; the dashed line indicates the resolution limit for the $4096^3$ grid.
The smallest grid resolution is reached at $t\approx 1.85$ for the A flow and at $t\approx 2.35$  for the C flow.
{\it Second and fourth panels:} Spectral index of the total energy, with values that  seem to settle  below $n=3$ for both configurations.
}  \label{fig:other1}
\end{center}
\end{figure}

\begin{figure}
 \begin{center}
 \includegraphics[width=.9\columnwidth]{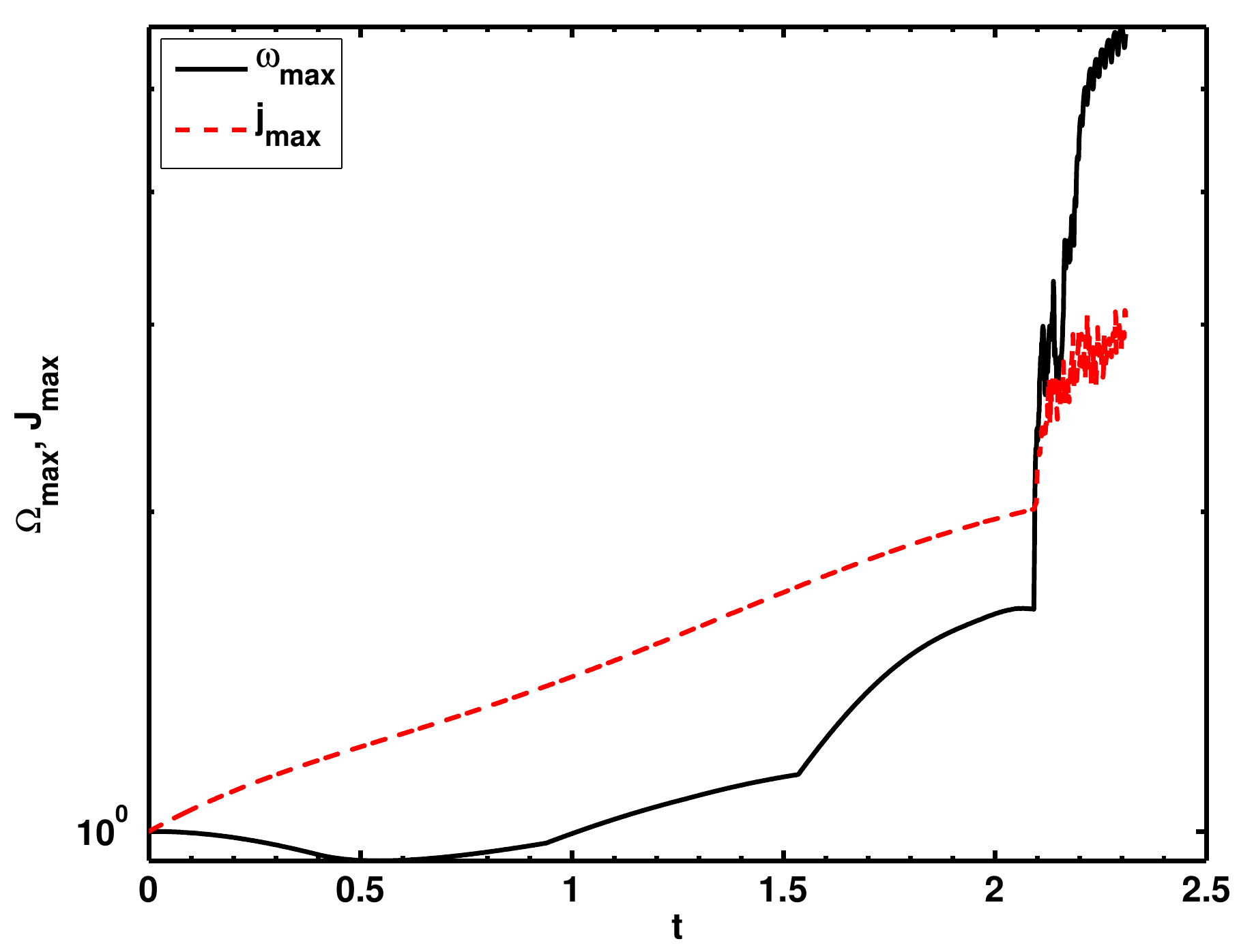}
 \includegraphics[width=.9\columnwidth ]{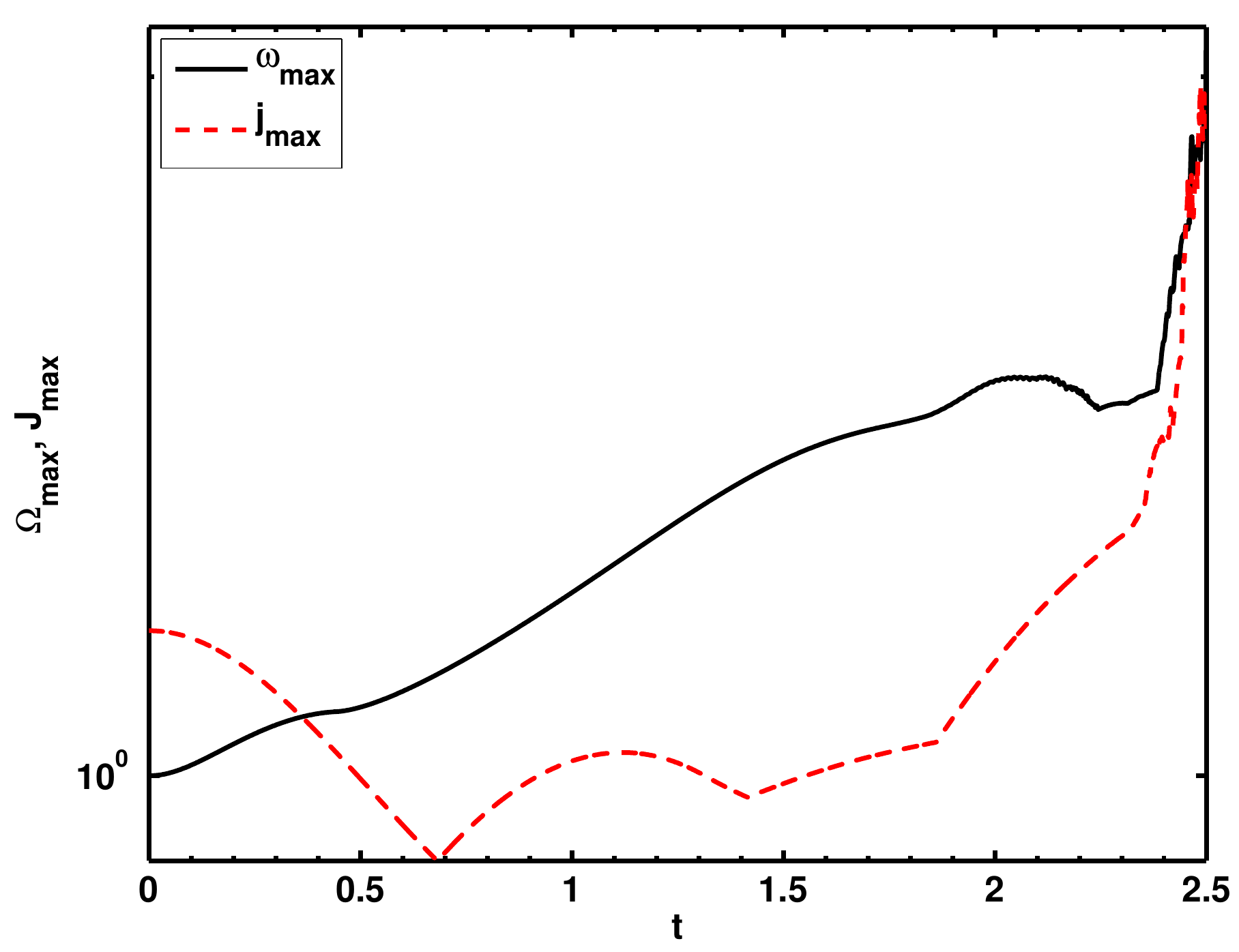}
 \includegraphics[width=.9\columnwidth ]{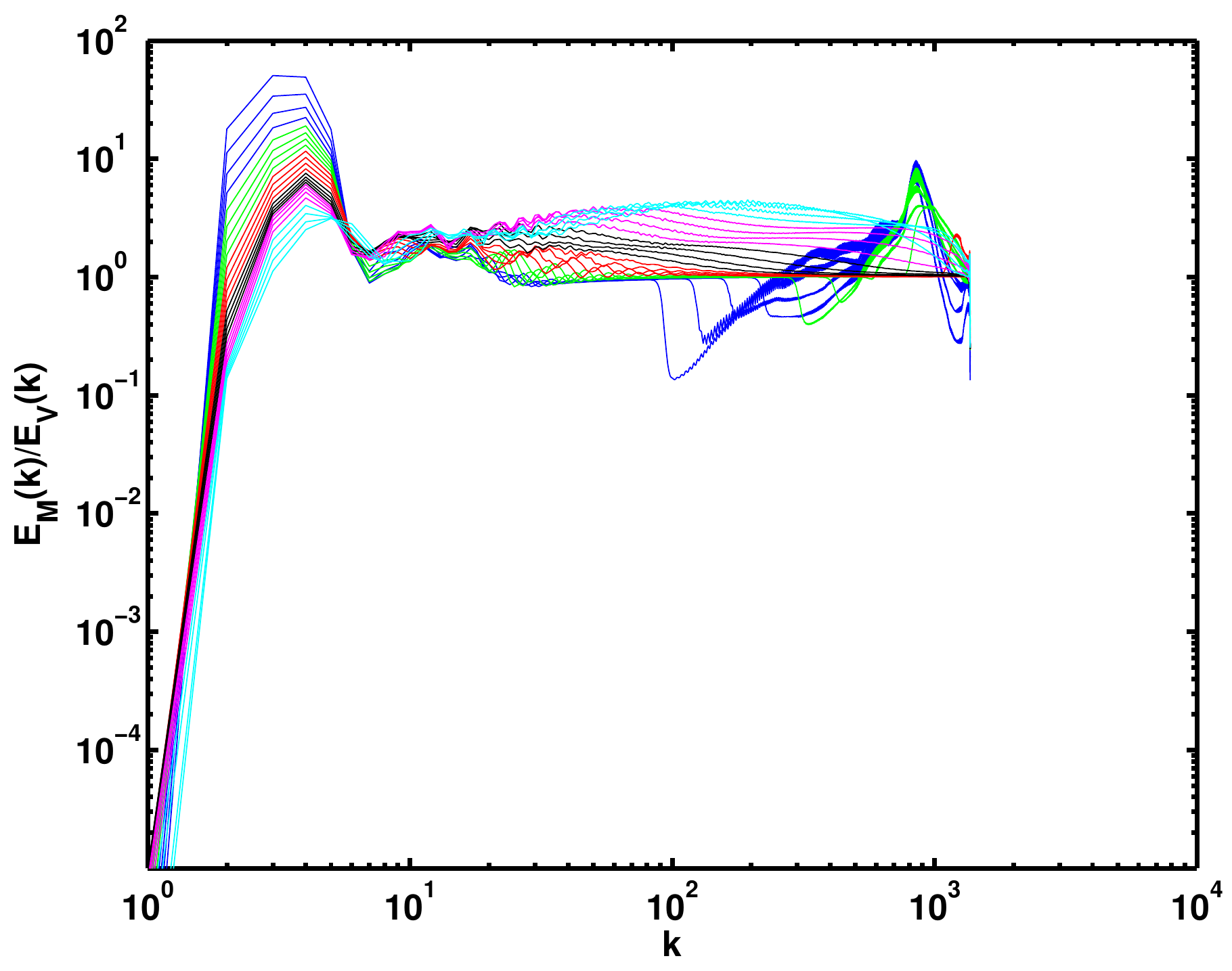}
 \caption{(color online)
  {\it Top, Center:}
Maxima of current (dashed, red) and vorticity (solid, black) over time for the A configuration ({\it top}), and for the C configuration ({\it center}). The abrupt changes at intermediate times are linked to the fact that the maxima jump from structure to structure. Note the rapid increase at $t> 1.85$ in the C flow and, and $t>2.35$ for the A flow, and the noise at late times due to the fact that small-scales have become insufficiently resolved.
{\it Bottom:}
Ratio of the magnetic to kinetic energy spectra, for the C configuration at different times, with dark blue ($0.43 \le t < 0.79$), green up to $t=1.15$, red up to $t=1.51$, black up to $t=1.87$, magenta up to $t=2.23$, and cyan up to $t=2.48$. Note the quasi-equipartition in the inertial range all the way to the cut-off, as well as  the excess magnetic energy at large scales.
}  \label{fig:other2}
  \end{center}
  \end{figure}

Finally, let us mention briefly how the two other initial conditions satisfying the TG symmetries behave, with ideal runs computed on grids of up to $4096^3$ points. Similar temporal evolutions seem to occur for both flows, as shown in Figs. \ref{fig:other1} and \ref{fig:other2}. The spectral indices seem to reach values smaller than in the Euler (ideal fluid) case for all configurations examined in this paper, systematically below a value $\approx 3$, with some oscillations in the conducting case (C flow, Fig.~\ref{fig:other1}, two lower panels).

The maxima of current and vorticity are displayed in Fig. \ref{fig:other2} for the A and C configurations. The C flow current maximum (Fig. \ref{fig:other2}, center) undergoes first a jump from structure to structure at relatively early times, followed by a traditional exponential phase corresponding to the thinning of current. This is followed again by a short and rapid further increase in the maximum which appears difficult to analyze in more detail, due to the fact that the  temporal interval during which this latest acceleration occurs is again too short, as for the insulating configuration analyzed in the previous sections. The vorticity maximum in the A flow shows a rather monotonic increase until the final acceleration in current. For the A flow (Fig. \ref{fig:other2} top) the monotonicity of the current and vorticity maxima are reversed compared to the C flow. We note that the temporal evolution of the integrated square vorticity and current for the A and C runs indicate that they become nearly equal for late times (not shown). This is due to the fact that, after the grid resolution is reached by the velocity and magnetic field structures, the evolution is that of a truncated system of Fourier modes which evolve, in the simplest case, to equipartition due to statistical equilibrium, as analyzed in \cite{frisch_75}; this begins to occur at late times in these computations, at a faster rate the smaller the scale.

Another instance of quasi-equipartition between kinetic and magnetic energy is occurring at earlier times, and is reminiscent of  what is observed in the dissipative case for many configurations (see, e.g., \cite{lee2, julia}).
Indeed, we see that the ratio of the spectra of magnetic and kinetic energy given in Fig. \ref{fig:other2} (bottom) for the C configuration is close to (and slightly above)  unity  from $k\approx 4$ up to the maximum wavenumber. This is also observed for the other two configurations examined in this paper and is consistent with the expression for the spectra obtained for ideal dynamics of a truncated system with zero (or negligible) helicity \cite{frisch_75}, in which case equipartition obtains. This appears to be another example where the ideal dynamics is consistent (and can be viewed as predictive of) dissipative (and/or forced) inertial range dynamics, as first clearly showed using direct numerical simulations in the fluid case in \cite{cicho_05a}. We  also note that such a quasi-equipartition of kinetic and magnetic energy, with in most cases a slight excess of the latter, has been observed for a long time in Solar Wind data \cite{goldstein}, and confirmed later by more detailed observations as well.

\section{Summary of results and conclusions}\label{s:conclu}

We have shown in this paper several new results concerning the ideal dynamics of MHD configurations, namely that:
{\bf  (i)} by increasing the resolution by a factor of three from our previous study, we still reproduce the results obtained in a $2048^3$ simulation up to the last time computed in that run, including an acceleration in the maximum of current and vorticity and in the decrease of the logarithmic decrement;
{\bf (ii)} in the new high-resolution simulation, we see yet a second acceleration of the formation of small scales at a later time, in a situation that is as well resolved as the previous acceleration was in the $2048^3$ simulation;
{\bf  (iii)} these two accelerations are clearly associated with changes in the structures in physical space of the current and vorticity;
{\bf  (iv)} these changes also pollute the small-scale spectrum, creating a limitation in practice to the applicability of the analyticity strip method;
{\bf   (v)} a new method, bridging the analyticity strip method and the so-called BKM criteria by means of sharp analysis inequalities, is extended to MHD, and allows us to rule out spurious singularities;
 {\bf   (vi)} the new method cannot completely rule out the existence of a finite-time singularity at a time between $t=2.33$ and $t=2.7$;
{\bf (vii)} the structures that seem to create this acceleration in the formation of small scales are related to the near collision and further spatial co-location of two current  sheets, and similarly for the vorticity;
{\bf (viii)} these results do not seem to be occurring only for one flow, but seem to take place as well in the other two configurations studied in this paper, up to equivalent resolutions of $4096^3$ points;
and finally
{\bf (ix)} a simple re-gridding technique, which allows for substantial savings in computer time, is shown to be entirely reliable provided a conservative threshold for applying the method is utilized.
We should note that in one case (that of the I configuration), the scale separation reached in the computation is unprecedented up to this point in time.

In summary, we have found that at high resolution, the most intense structures that develop in ideal MHD come from the near collision and later from the near juxtaposition of two current and vorticity sheets. The maxima of these small-scale fields undergo abrupt jumps twice, and it will be necessary to pursue this computation at yet higher resolutions to see whether the criteria for a singularity to develop or not are satisfied, by monitoring for a time that is sufficiently long the maxima of current and vorticity and to compute other diagnostics as well.

We remark that our pursuit is not just a brute-force increase in resolution. In fact, we have made use of a new analytical tool, that bridges two known singularity criteria (BKM-type theorem for MHD and analyticity-strip method), leading to a new method for ruling out spurious indications of singularity. We have applied this method to the current configuration under study at the highest resolution achieved in this paper, and concluded that the existence of a finite-time singularity at a time between $t=2.33$ and $t=2.7$ cannot be completely ruled out. While it would be desirable to produce a more specific statement in this regard, there is one fact that makes it difficult to advance further: the values of $\|\bomega(\cdot,t)\|_{\infty}$ and $\|\mathbf{j}(\cdot,t)\|_{\infty}$, needed for testing singular behavior in the framework of the BKM theorem are currently measured using collocation-point data, a standard procedure that leads to ``noise'' or error in the data. This noise is evident as tiny oscillations in Fig.~\ref{fig:om_j_max}(b) which add an uncertainty to the computation of slopes. The source of this noise was discovered recently in \cite{bustamante2011}, in the context of the more controllable inviscid Burgers one-dimensional flow. There, as in our MHD case, the systematic periodic sampling of collocation-point maxima introduces an error in the precision of the measurement with respect to the true value of maxima. The error oscillates in time; its frequency grows with the numerical resolution used if the time step is determined by a fixed-ratio CFL condition. Moreover, the amplitude of the error depends on the spatial profile of the maximum computed, so that the error increases as the structures become more peaked. In \cite{bustamante2011} the solution to this problem was proposed and has two levels of complexity: at the simplest level, a post-processing computation of extrapolated values of the maxima of vorticity and current can eliminate partially the oscillatory part of the error. At the deepest level, the application of an adaptive time stepping beyond CFL, so that the product $\Delta t \times \left(\|\bomega(\cdot,t)\|_{\infty}+\|\mathbf{j}(\cdot,t)\|_{\infty}\right)$ remains constant, can improve the precision in the computation of vorticity and current maxima by a factor of $10^2,$ at no extra cost in computational time and memory  \cite{bustamante2011}. In our future work we will implement these procedures so we can have more robust evidence regarding the hypothesis of finite-time singularity in MHD. 

One dynamical effect that can play a role in stopping a putative singularity is the phenomenon of dynamic alignment that is rather ubiquitous in turbulent flows. For example, it was shown in  \cite{matthaeus_08} that  the alignment of vorticity with shear or pressure gradients, and equivalently of magnetic field and shear, enhances point-wise helicity (kinetic helicity in the former case, cross helicity in the second case), although the global norms are conserved, and it does so in a time of the order of the eddy turn-over time. In fact, an alignment between all variables involved in the nonlinear terms of MHD, namely, velocity and magnetic field in Ohm's law \cite{meneg}, velocity and vorticity in the Lamb vector, and current and magnetic field in the Lorentz force,  occurs rather systematically, in particular the latter \cite{servidio}. It is not clear what the effect of dissipation is in these alignment properties, or whether such alignment tendencies would be sufficient to prevent singularities to occur in the ideal case. In that light, a more detailed analysis of the local properties of the flow in the vicinity of the current and vorticity maxima will be undertaken in a follow-up paper.
Furthermore, ideal and dissipative flows have common properties because of their nonlinear multi-scale interactions. The lack of universality, found in decaying flows with imposed Taylor-Green symmetries \cite{lee2} is also found in the forced case \cite{krstu_12}, and thus it is an open problem to see whether it will  occur in the ideal case, although the differences between inertial indices is small and thus requires high resolutions and long-time integration.

A theory of turbulent flows is still lacking, and yet such flows are ubiquitous in nature and are an integral part of the problem of weather prediction, of climate assessment, of understanding the formation and prediction of extreme events such as tornadoes and hurricanes, of reconnection events in space physics such as solar flares and coronal mass ejections, plasmoids, and in disruptive plasmas. Such flows develop intense small scale structures in the form of vortex and current sheets and filaments with power-law scaling properties and departure from Gaussianity attributed to  intermittency.
 Similarly in the ideal case at intermediate times and intermediate scales, a classical turbulent spectrum has been observed recently for fluids \cite{cicho_05a, krstu_09}, with at smaller scales the statistical equilibrium that can be derived analytically using the quadratic invariants preserved by the truncation (see \cite{frisch_75} for 3D MHD), the whole flow evolving  progressively towards flux-less Gaussian equilibrium solutions.
What is lacking is, among other things, a statistical description of the small scales, and a prediction of long-time large-scale dynamics with ensuing modified transport properties. By combining this study with a well-resolved high Reynolds number dissipative run, one may be able to establish in 3D-MHD the link between the role of ideal nonlinear dynamics, and dissipative-induced reconnection (see e.g. \cite{kerr_89}), leading to finite dissipation in the limit of zero viscosity and magnetic resistivity  as shown in both two-dimensional \cite{biskamp, pp89} and three-dimensional cases \cite{mininni_09}. This may shed light on dissipation processes in turbulent conducting flows, and on the role of non-local interactions between disparate scales \cite{shell_I, shell_II} in MHD when compared to the Euler (fluid) case (see also \cite{gomez}), thus leading to better estimations of the energy dissipation rate controlled by turbulence in astrophysics and space physics.

\begin{acknowledgments}
 The National Center for Atmospheric Research is sponsored by NSF.
 Computations were performed on the OLCF ``jaguar'' system (Oak Ridge),
  and computer time was provided through a 2011 INCITE allocation, number 16013.
  Marc Brachet acknowledges a travel fund grant from the NCAR Geophysical Turbulence Program. Miguel Bustamante acknowledges the support of UCD Seed Funding SF564.
\end{acknowledgments}


\begin{thebibliography}{23}

\bibitem{constantin1995}
 Constantin P., Procaccia I. and  Segel D.,  {\it Phys. Rev. E} {\bf 51}, 3207, 1995

\bibitem{eyink_euler}
 Eyink G. {\it et al.}, 
 {\it Physica D} {\bf 237}, {\it  ``General Introduction, Euler Equations: 250 Years on (EE250)''}, 2008

\bibitem{vieille_82}
 Vieillefosse P., {\it J. Phys.} {\bf 43}, 837, 1982

\bibitem{vieille_84}
 Vieillefosse P., {\it Physica A} {\bf 125}, 150, 1984

\bibitem{klapper_96}
 Klapper I.,  A. Rado and M. Tabor, {\it Phys. Plasm.} {\bf 3}, 4281, 1996 

\bibitem{BKM}
Beale J.T., T. Kato and A. Majda,  
{\it Commun. Math. Phys.} {\bf 94},  61, 1984

\bibitem{kerr_05}
Kerr R.,  
{\it Phys. Fluids} {\bf 17}, 075103, 2005

\bibitem{gibbon_08}
 Gibbon J.D.,  {\it Physica D} {\bf 237},1894, 2008

\bibitem{branden1995}
Brandenburg A., Procaccia I. and  Segel D.,  {\it Phys. Plasmas} {\bf 2}, 1148, 1995

\bibitem{pouquet_sanminiato_96}
  Pouquet A., {\it Lecture Notes in Physics} (Springer)  
{\bf 468}, 163, 1996 

\bibitem{brachet_83}
Brachet M.E. {\it et al.},
J. Fluid Mech. {\bf 130}, 411, 1983

\bibitem{kida_85}
 Kida S.,  {\it J. Phys. Soc. Japan.} {\bf 54}, 2132, 1985

 \bibitem{boratav_94}
Boratav O.N. and R. B. Pelz,  {\it Phys. Fluids} {\bf 6}, 2757, 1994

\bibitem{kerr_89}
Kerr R. and F. Hussain,  
 {\it Physica D} {\bf 37}, 474, 1989

 \bibitem{kerr_93}
 Kerr  R.M.,  
{\it Phys. Fluids} {\bf 5}, 1725, 1993

\bibitem{pelz_01}
 Pelz R., {\it J. Fluid Mech.} {\bf 444}, 299, 2001 

\bibitem{hou_06}
Hou T.Y. and  R. Li,  
{\it J. Nonlin. Sci.} {\bf 16}, 639, 2006

\bibitem{hou_08}
Hou T.Y. and  R. Li,  {\it Physica D} {\bf 237}, 1937, 2008

\bibitem{kerr_08}
 Bustamante M. D. and Kerr R.,   {\it Physica D} {\bf 237}, 1912, 2008

\bibitem{bustamante}
 Bustamante M. D. and M.E. Brachet, ``On the interplay between the BKM theorem and the analyticity-strip method to investigate numerically the incompressible Euler singularity problem,'' {\it Phys. Rev. E}, in press (2011).

\bibitem{caflisch}
Caflisch R.E., Klapper I. and Steele G.,  {\it Comm. Math. Phys.} {\bf 184}, 443, 1997 

\bibitem{uf_OT2D}
Frisch U. {\it et al.},  
{\it J. M\'ec. Th\'eor. Appl.} {\bf 2}, 191, 1983

\bibitem{klapper_93}
 Klapper I. and M. Tabor, {\it Geophys. Astrophys. Fluid Dyn.} {\bf 73}, 109, 1993 

\bibitem{grauer_97}
 Grauer R. and C. Mariani, 
{\it Phys. Rev. Lett.} {\bf 84}, 4850, 1997 -

\bibitem{grauer_98a}
Grauer R. and Mariani C.,  {\it Phys. Plasmas} {\bf 5}, 2544, 1998 

\bibitem{klapper_98}
 Klapper I., {\it Phys. Plasm.} {\bf 5}, 910, 1998 

 \bibitem{krstu_11}
  Krstulovic, G., M.E. Brachet and A. Pouquet,
 {\it Phys. Rev. E} {\bf 84}, 016410, 2011 

\bibitem{kerr_branden_99}
Kerr R. and Brandenburg A.,  {\it Phys. Rev. Lett.} {\bf 83}, 1155, 1999

\bibitem{grauer_98b}
 Grauer R., C. Marliani and K. Germaschewski,  
 {\it Phys. Rev. Lett.} {\bf 80},  4177, 1998

 \bibitem{grauer_00}
  Grauer R. and Mariani C.,  {\it Phys. Rev. Lett.} {\bf 84}, 4850, 2000

\bibitem{grafke}
Grafke T. {\it et al.},  {\it Physica D} {\bf 237}, 1932, 2008 

\bibitem{lee_ideal}
 Lee E. {\it et al.},  
{\it Phys. Rev. E} {\bf 78}, 066401, 2008

\bibitem{lee2}
Lee E. {\it et al.}, 
{\it Phys. Rev. E} {\bf 81}, 016318, 2010

\bibitem{veltri1999}
Veltri P.L., 
{\it Plasma Phys. Control. Fusion} {\bf 41}, A787, 1999 

\bibitem{veltri2009}
 Veltri P.L. {\it et al.}, in {\sl Encyclopedia of Complexity and System Science} {\bf19}, R.A. Meyers Ed., Springer, 2009

\bibitem{veltri2005}
 Veltri P.L. {\it et al.}, 
{\it Nonlinear Processes in Geophysics}, {\bf 12},  245, 2005 

\bibitem{Lin2009}
Lin C.C. {\it et al.},  {\it J. Geophys. Res.} {\bf 114}, A08102, 2009

\bibitem{woltjer}
 Woltjer L., {\it Proc. Nat. Acad. Sci.} {\bf 44}, 833, 1958

\bibitem{brachet_92}
Brachet M.E. et al.,  {\it Phys. Fluids} {\bf A 4}, 284, 1992
f
\bibitem{cicho_05a}
Cichowlas C. {\it et al.}, 
{\it Phys. Rev. Lett.} {\bf 95}, 264502, 2005

\bibitem{gafd}
 Pouquet A. {\it et al.}, 
 {\it Geophys. Astrophys. Fluid Dyn.}, {\bf 104}, 115, 2010 

\bibitem{julia}
J. Stawarz, A. Pouquet and M-E. Brachet, 
 {\it Phys. Rev. E} {\bf 86}, 036307 (2012)

   \bibitem{matthaeus_08}
Matthaeus W.H. {\it et al.},  
 {\it Phys. Rev. Lett.}{\bf 100}, 085003, 2008

\bibitem{bardos_82}
Bardos C., in {\sl Nonlinear problems: Present and future}, Bishop A. et al. Eds.,
North-Holland, 1982

\bibitem{galtier_00}
Galtier S.  {\it et al.}, 
{ J. Plasma Phys.} {\bf 63}, 447, 2000

\bibitem{galtier_02}
Galtier S.  {\it et al.},  
{ Astrophys. J.  Lett.} {\bf 564}, L49, 2002 

\bibitem{gottlieb}
Gottlieb D. and S. A. Orszag,  {\sl Numerical Analysis of Spectral Methods: Theory and Application}. SIAM, Philadelphia, 1977

\bibitem{hybrid}
 Mininni P. D. {\it et al.}, 
 {\it Parallel Computing}, {\bf 37}, 316, 2011 

\bibitem{yeung}
Yeung P.K {\it et al.},  {\it Phys. Fluids} {\bf 17}, 081703; and Donzis et al. 2008 {TeraGrid Conf.}, Las Vegas, NV., 2005

\bibitem{gaspar}
Rosenberg D. {\it et al.},  
 {\it J. Comp. Phys.} {\bf 215} 59, 2006

 \bibitem{NJP}
Rosenberg D., Pouquet, A. and  Mininni P.,
 {\it New J. Phys.}, {\bf 9}, 304, 2007

\bibitem{sun}
Sun J. {\it et al.}, 
{\it J. Atmos. Sci.} {\bf 69}, 338, 2011 

\bibitem{bardos_07}
 Bardos,C. and E. Titi, 
{\it Russian Math. Surveys} {\bf 62}, 409, 2007

\bibitem{cicho_05b} Cichowlas C. and Brachet M.E.,  {\it Fluid Dyn. Res.} {\bf 36}, 239, 2005 

\bibitem{vapor}
 Clyne J. {\it et al.},  
 New J. Phys. {\bf 9}, 301, 2007

\bibitem{mininni_08}
Mininni P.D. {\it et al.},  
 {\it New J. Phys.}  {\bf 10}, 125007, 2008

\bibitem{hasegawa_04}
 Hasegawa H. {\it et al.}, 
{\it Nature} {\bf 430}, 755, 2004 

   \bibitem{phan_06}
 Phan T.D. {\it et al.},
{  Nature} {\bf 439}, 175, 2006

\bibitem{mininni_06a}
Mininni P. D., Alexakis A. and Pouquet A.,  
{\it Phys. Rev. E} {\bf  74}, 016303, 2006

\bibitem{mininni_06b}
Mininni P.D. and Pouquet A.,    {\it Phys. Rev. Lett.} {\bf 97}, 244503, 2006

\bibitem{mininni_09}
  Mininni P.D. and Pouquet A.,  
 {\it Phys. Rev. E}  {\bf 80}, 025401, 2009

 \bibitem{constantin1994}
Constantin P., 
{\it SIAM Rev.}, {\bf 36}, 73, 1994  

  \bibitem{constantin2008}
Constantin P., 
{\it Physica D Rev.}, {\bf 237}, 1926, 2008  

\bibitem{frisch_75}
 Frisch U. {\it et al.}, 
 {J. Fluid Mech.}, {\bf 68}, 769, 1975 

\bibitem{goldstein}  Matthaeus W.H. and M. Goldstein, {\it J. Geophys. Res.} {\bf 87}, 6011, 1982 

\bibitem{bustamante2011}
 Bustamante M. D., {\it Physica D} {\bf 240}, 1092, 2011

\bibitem{meneg}
 Meneguzzi M. {\it et al.}, {\it J. Comput. Phys.} {\bf 123}, 32, 1996

\bibitem{servidio}
 Servidio S., W. H. Matthaeus and P. Dmitruk, 
{\it Phys. Rev. Lett.} {\bf 100}, 095005, 2008

\bibitem{krstu_12}
Krstulovic G., M.E. Brachet and A. Pouquet,
``Forced dynamics of three-dimensional MHD flows implementing  the Taylor-Green symmetries,''
in preparation, 2012 

 \bibitem{krstu_09} Krstulovic G. {\it et al.}, 
 {\it Phys. Rev. E} {\bf 79}, 056304, 2009 

\bibitem{biskamp}
 Biskamp D., {\sl Nonlinear Magnetohydrodynamics,} Cambridge University Press, 1993

\bibitem{pp89}
Politano H., Pouquet, A. and Sulem, P.L., 
{\it Physics Fluids B} {\bf 1}, 2330, 1989 

\bibitem{shell_I}
 Alexakis A., Mininni P.D., Pouquet A.,  {\it Phys. Rev. E} {\bf  72}, 046301, 2005

\bibitem{shell_II}
 Mininni P.D., A. Alexakis and A. Pouquet,
 {\it Phys. Rev. E} {\bf 72}, 046302, 2005

\bibitem{gomez}
 Gomez T., H. Politano and A. Pouquet,
{\it Phys. Fluids}, {\bf 11}, 2298, 1999 

\end{thebibliography}
\end{document}